\documentclass{PoS}

\usepackage[utf8x]{inputenc}
\usepackage{amsmath}
\usepackage{graphicx,caption}
\usepackage{float}
\usepackage[caption=false]{subfig}
\usepackage{tikz}
\usetikzlibrary{snakes}
\usepackage{epstopdf}
\usepackage{slashed}

\title{Impact of the Delta (1232) resonance on neutral pion photoproduction in chiral perturbation theory.}

\ShortTitle{Impact of the Delta (1232) resonance in neutral pion photoproduction in chiral perturbation theory.}

\author{\speaker{Lloyd Cawthorne}\\%
        The University of Manchester, Oxford Rd, Manchester M13 9PL, United Kingdom\\
        E-mail: \email{lloyd.cawthorne@postgrad.manchester.ac.uk}}

\author{Judith McGovern\\
        The University of Manchester, Oxford Rd, Manchester M13 9PL, United Kingdom\\
        E-mail: \email{judith.mcgovern@manchester.ac.uk}}

\abstract{We present an ongoing project to assess the importance of D-waves and the $\Delta (1232)$ resonance for descriptions of neutral pion photoproduction in Heavy Baryon Chiral Perturbation Theory. This research has been motivated by data published by the A2 and CB-TAPS collaborations at MAMI \cite{hornidge13}. This data has reached unprecedented levels of accuracy from threshold through to the $\Delta$ resonance. Accompanying the experimental work, there has also been a series of publications studying the theory that show that, to go beyond an energy of $E_\gamma=170$ MeV, it is necessary to include other aspects, in particular the $\Delta (1232)$ as a degree of freedom \cite{blin15} and possibly higher partial waves \cite{fernandez-ramirez09}.}

\FullConference{The 8th International Workshop on Chiral Dynamics, CD2015 ***\\
		29 June 2015 - 03 July 2015\\
		Pisa,Italy}

\begin{document}

\section{Introduction}
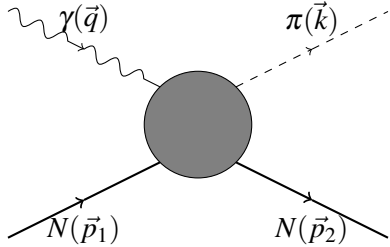
\begin{figure}
\begin{center}
\begin{tikzpicture}[domain=-2.5:12,scale=1]

     \draw[->,thick] (-5,-1) -- (-4,-0.5) node[shift={(0,-0.4)}] {$N(\vec{p_1})$};
    \draw[-,thick] (-4,-0.5) -- (-3,0){};
      \draw[->,thick] (-2,0) -- (-1,-0.5) node[shift={(0,-0.4)}] {$N(\vec{p_2})$};
        \draw[-,thick] (-1,-0.5) -- (0,-1){};
	\node[draw,circle,inner sep= 0.5cm,fill=gray] at (-2.5,0.5){};
      \draw[->,dashed] (-2,1) -- (-1,1.5) node[shift={(0,0.4)}] {$\pi(\vec{k})$};
      \draw[-,dashed] (-1,1.5) -- (0,2) {};
      \draw[->,snake=coil,segment aspect =0] (-5,2) -- (-4,1.5) node[shift={(0,0.4)}] {$\gamma(\vec{q})$};
      \draw[-,snake=coil,segment aspect =0] (-4,1.5) -- (-3,1){};
             
\end{tikzpicture}
\caption{Diagram of the interaction, the grey circle represents all intermediate states.} \label{fig:pict}
\end{center}
\end{figure}

The goal of this work is to use chiral effective field theory (EFT) to describe the process $p+\gamma \rightarrow p+\pi^0$ from threshold to energies approaching the $\Delta (1232)$ resonance. The reaction is illustrated in figure \ref{fig:pict}.

For over two decades pion photoproduction has been studied using Chiral Perturbation Theory ($\chi$PT). Bernard {\it et al.\ }were the first to do so, working with the $\mathcal{O}(p^3)$ relativistic theory \cite{bernard92} to describe the data from Mainz \cite{mazzucato86} and Saclay \cite{beck90}. The theory successfully described the experimental data from threshold, $E_\gamma \approx145$ MeV, to $E_\gamma \approx 160$ MeV. After the formulation of Heavy Baryon Chiral Perturbation Theory (HB$\chi$PT), they revisited the subject extending the calculation to $\mathcal{O}(p^4)$ \cite{bernard96,bernard01}. The authors also used the HB method to accurately describe cusp effects, in particular in the $E_{0+}$ multipole. This work improved on previous results by fitting the data up to $E_\gamma\approx 165$ MeV.

Following the publication of the results from the A2 and CB-TAPS collaborations at the Mainz Microtron (MAMI) there has been a renewed interest in pion photoproduction \cite{hornidge13}. This data was first analysed by Fern{\'a}ndez-Ram{\'i}rez {\it et al.} in the fourth-order heavy baryon approach \cite{fernandez-ramirez13}. Separately, Hilt {\it et al.} used the $\mathcal{O}(p^4)$ relativistic Extended On Mass Shell (EOMS) calculation and found that the HB approach performed slightly better \cite{hilt13}. They concluded that both theories fail to describe the process beyond energies of $E_\gamma \approx 170$ MeV. Two explanations have been suggested for these failures: the need for the $\Delta (1232)$ to be included as a degree of freedom; and the possibility that D-waves (or higher) have a greater impact on observables than expected \cite{fernandez-ramirez09}. It should not come as a surprise that the $\Delta(1232)$ is required when studying photoproduction well beyond threshold, as the mass difference between it and the nucleon is not large, $\Delta_M=m_\Delta-m_N\approx 290$ MeV. It has already been shown by Blin {\it et al.} that the inclusion of the $\Delta$ in a $\mathcal{O}(p^3)$ EOMS theory substantially improves the fit \cite{blin15}. 

In this article we will concentrate on the effects of including the $\Delta$ in the $\mathcal{O}(p^4)$ HB$\chi$PT framework of Bernard {\it et al.} \cite{bernard96,bernard01}. This is done by using the ``little $\delta$'' scheme as developed by Pascalutsa and Phillips \cite{pascalutsa03}. We also attempt to study the effects of D-waves on the observables.

\section{Formal aspects}

\subsection{Framework}

The A2 and CB-TAPS collaborations have produced data on the differential cross section and the photon asymmetry \cite{hornidge13}. Both observables can be expressed through the scattering amplitude; details can be found in the theory-independent study by Hanstein {\it et al.} \cite{hanstein98}. 

\begin{figure}
\centering
\subfloat[]{\includegraphics[width=2.8cm]{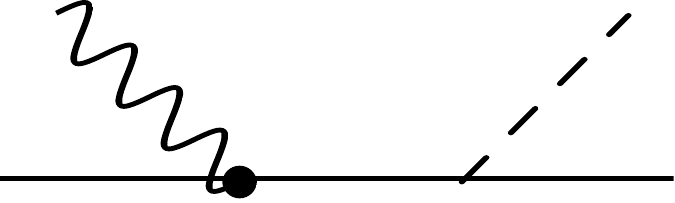}}\;
\subfloat[]{\includegraphics[width=2.8cm]{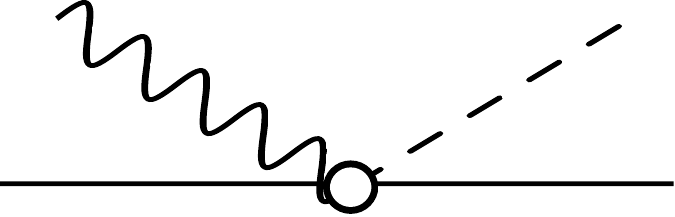}}\;
\subfloat[]{\includegraphics[width=2.8cm]{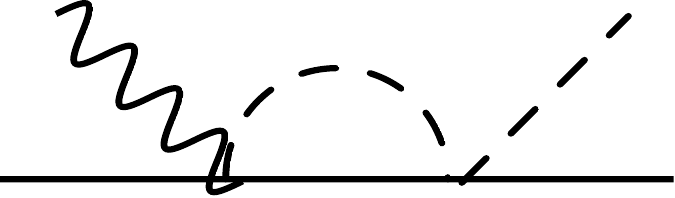}}\\
\subfloat[]{\includegraphics[width=2.8cm]{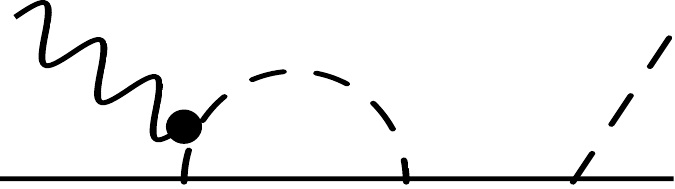}}\;
\subfloat[]{\includegraphics[width=2.8cm]{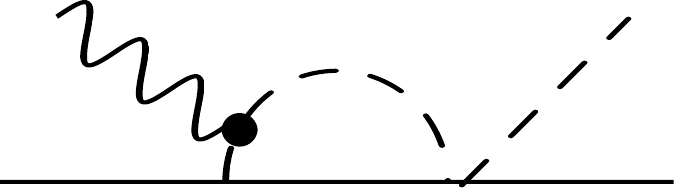}}\;
\subfloat[]{\includegraphics[width=2.8cm]{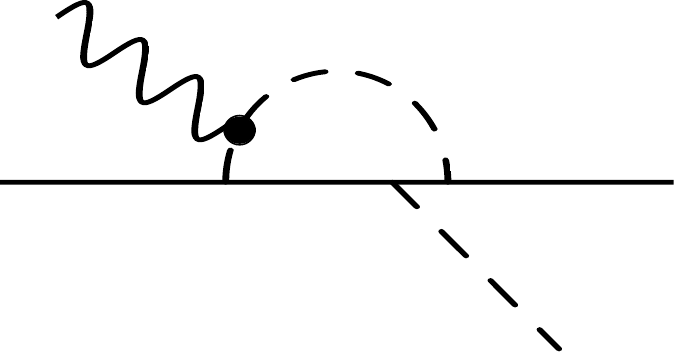}}

\caption{Feynman diagrams contributing to neutral pion photoproduction up to $\mathcal{O}(p^3)$. Filled circles represent second order vertices, open circles represent vertices of order one through to three. Diagram (a) is of $\mathcal{O}(p^2)$, the third-order graphs with the same topology are not shown. We have omitted the different orderings of diagrams (a), (c), (d) and (e) for brevity.}

\label{F:pin3}

\end{figure}

\begin{figure}
\centering
\subfloat[]{\includegraphics[width=2.8cm]{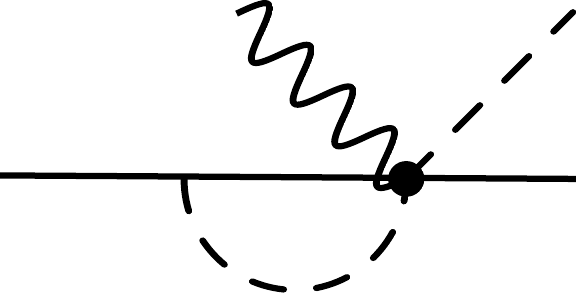}}\;
\subfloat[]{\includegraphics[width=2.8cm]{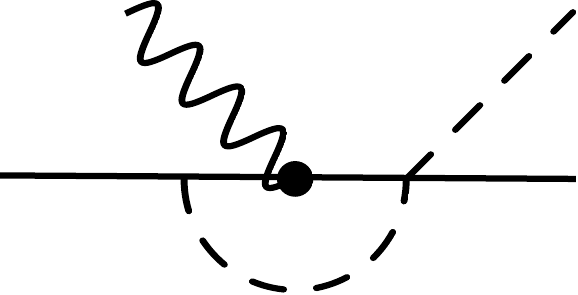}}\;
\subfloat[]{\includegraphics[width=2.8cm]{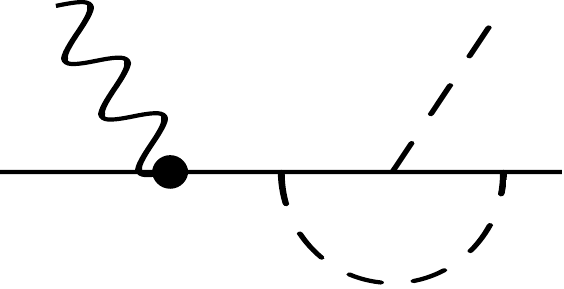}}\;
\subfloat[]{\includegraphics[width=2.8cm]{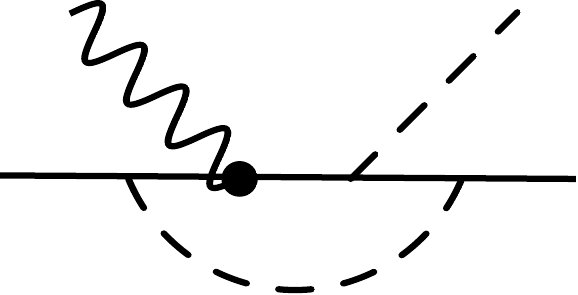}}\\
\subfloat[]{\includegraphics[width=2.8cm]{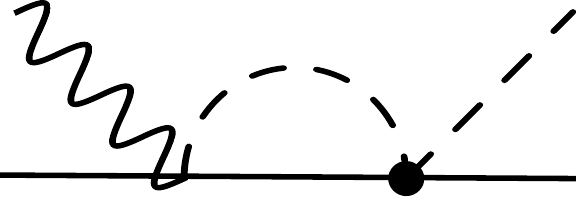}}\;
\subfloat[]{\includegraphics[width=2.8cm]{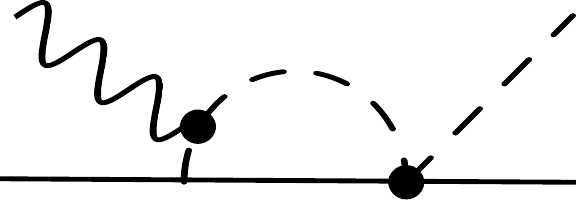}}

\caption{$\mathcal{O}(p^4)$ Feynman diagrams contributing to neutral pion photoproduction that introduce new topologies or LECs. Filled circles represent second order vertices. We have omitted the different orderings of diagrams for brevity.}

\label{F:order4d}

\end{figure}

Due to the small mass difference between charged and neutral pions there are cusp effects in neutral pion photoproduction. To describe these effects in $\chi$PT, $\mathcal{O}(p^3)$ loop diagrams must be included that contain a virtual charged pion \cite{bernard92, bernard96}. These diagrams are shown in figure \ref{F:pin3} for the HB approach. Furthermore, at third-order, there are also sub-leading Born diagrams that involve low energy constants (LECs) that require fitting to data. We will discuss this in more detail in section \ref{fitprocedure}. The Lagrangian used can be found in \cite{fettes00}. The $\mathcal{O}(p^4)$ diagrams that have new topologies or LECs are shown in figure \ref{F:order4d}. The fourth-order diagrams that are not shown in figure \ref{F:order4d} will only introduce $1/m_N$ corrections or mass re-normalizations to the already shown third-order graphs.

\subsection{D-waves}

As has been stressed by Fern\'andez-Ram\'irez {\it et al.} D-waves may play an important role when analysing data at energies beyond the threshold region \cite{fernandez-ramirez09}. The idea is not that pure D-wave contributions to the amplitudes are significant, but that they might interfere with the dominant S- and P-waves. This would be seen in the data as deviation from a $\mathrm{cos}^2\theta$ dependence for $d\sigma/d\Omega$ or a $\mathrm{sin}^2\theta$ for $\Sigma$.
 
The HB work of Bernard {\it et al.} was conducted to study photoproduction up to 20 MeV above threshold \cite{bernard96,bernard01}; for that reason they truncate their amplitudes to S- and P-waves as higher partial waves do not have significant effects at those energies. As we are still in the process of re-calculating the fourth-order HB amplitudes, for now we have included waves with $L=2$ (and higher) by using the $\mathcal{O}(p^3)$ relativistic calculation of Bernard {\it et al.} \cite{bernard92}.

\subsection{$\Delta(1232)$}

We include the $\Delta(1232)$ state as a new degree of freedom using the ``little $\delta$'' counting, as described by Pascalutsa and Phillips  \cite{pascalutsa03}. We will take some time to outline this procedure as the power counting depends on the energy of the system; this means that some diagrams that are N$^2$LO in the threshold regime are promoted to NLO at energies close to the $\Delta(1232)$ resonance.

Close to threshold the pion energy will be roughly the same size as its mass, $\omega_\pi \approx M_\pi \approx 140$ MeV, and the chiral symmetry breaking scale is approximately the same size as the $\rho$ mass, $\Lambda_{SB} \approx 700$ MeV. So for energies close to threshold we have two small scales:

\begin{equation} 
p=\frac{\omega_\pi}{\Lambda_{SB}}\approx\frac{M_\pi}{\Lambda_{SB}}\approx 0.2, \qquad \delta = \frac{\Delta_M}{\Lambda_{SB}}\approx 0.4.
\end{equation}

We now have a choice as to how to proceed. We could either expand in each scale independently, or we could find a relation between the two scales and simply expand in terms of one of them. We have chosen the latter method to simplify our work, using the relation $p\approx \delta^2$. Examining the $\Delta$ propagator using the heavy baryon method (for schematic purposes only), and noting that $M_\pi \ll \Delta_M$:

\begin{equation} 
S_\Delta\left(\omega_\pi\sim M_\pi\right)\sim \frac{1}{\Delta_M \pm \omega_\pi},
\end{equation}
which scales as $p^{-1/2}$, or $\delta^{-1}$.

\begin{figure}
\centering
\subfloat{\includegraphics[width=2.8cm]{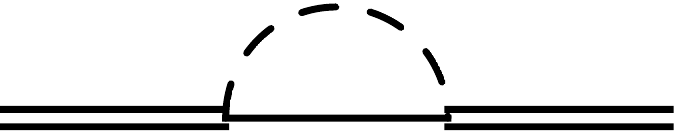}}
\caption{Virtual pion loop giving the self-energy of the $\Delta(1232)$.}

\label{F:selfenergy}

\end{figure}

However this is only valid at energies close to threshold. At energies near resonance $\left(\omega_\pi \approx \Delta_M\right)$ the relations outlined above change:

\begin{equation} 
p=\frac{\omega_\pi}{\Lambda_{SB}}\approx\frac{\Delta_M}{\Lambda_{SB}}=\delta.
\end{equation}
In this regime the $\Delta$ propagator also changes. Near its resonance we must take into account its self-energy, see fig. \ref{F:selfenergy}, which in turn gives it a width. The real part of the self-energy can be absorbed into the mass of the $\Delta$ or the wave-function re-normalisation, but the imaginary part is rapidly varying with energy and hence influences the observables.

Close to resonance the propagator takes the following form:

\begin{equation}
 S_\Delta(\omega_\pi\sim \Delta_M) \sim \frac{1}{\Delta_M + i \mathrm{Im}\left[ \Sigma_\Delta\right] - \omega_\pi},
\end{equation}
where

\begin{equation}
 \mathrm{Im}\left[\Sigma_\Delta(s)\right]=-\left(\frac{g_{\pi N\Delta}}{2 m_\Delta}\right)^2\frac{\left(\sqrt{s}+m_N\right)^2-M_\pi^2}{48 \pi m_\Delta^2} k_\pi^3.
\end{equation}
At resonance we are left with a purely imaginary propagator inversely proportional to the width. As Im$\left[\Sigma_\Delta\right]$ scales as $p^3$, the propagator scales as $p^{-3}\left(=\delta^{-3}\right)$ for $\sqrt{s}-m_N=\Delta_M$ $\left(\omega_\pi\approx\Delta_M\right)$. In other words, the effects of the $\Delta$ are promoted to LO as the power counting of its propagator changes from $p^{-1/2}$ to $p^{-3}$. This change in power counting will only occur for diagrams that are reducible across the $\Delta$ propagator, i.e. diagrams where $p_\Delta^2=s$.

\begin{figure}
\centering
\subfloat[]{\includegraphics[width=2.8cm]{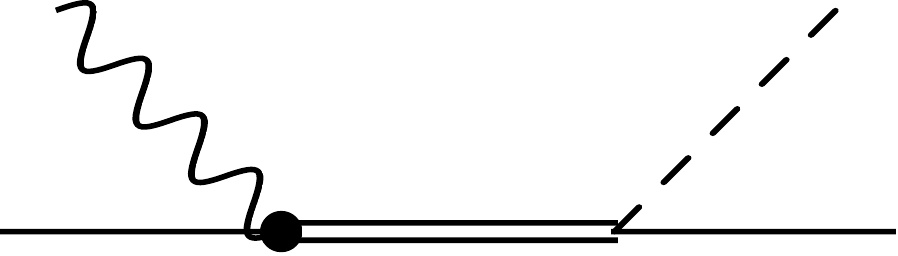}} \;
\subfloat[]{\includegraphics[width=2.8cm]{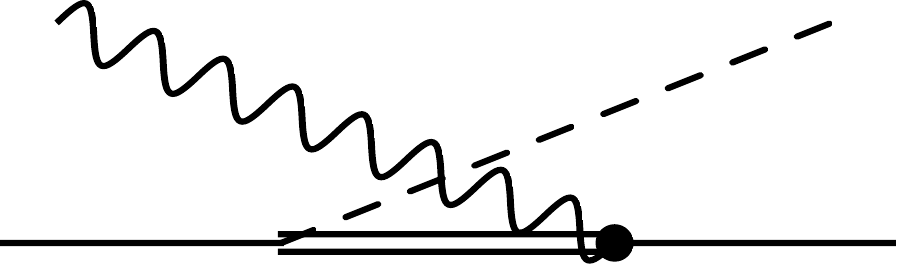}} \;
\subfloat[]{\includegraphics[width=2.8cm]{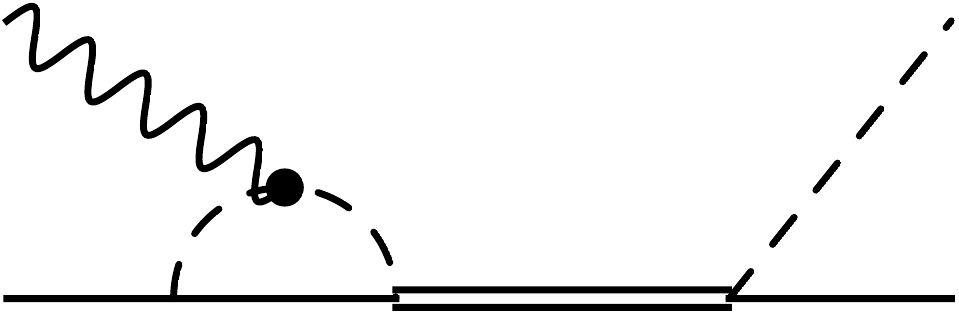}}\;
\subfloat[]{\includegraphics[width=2.8cm]{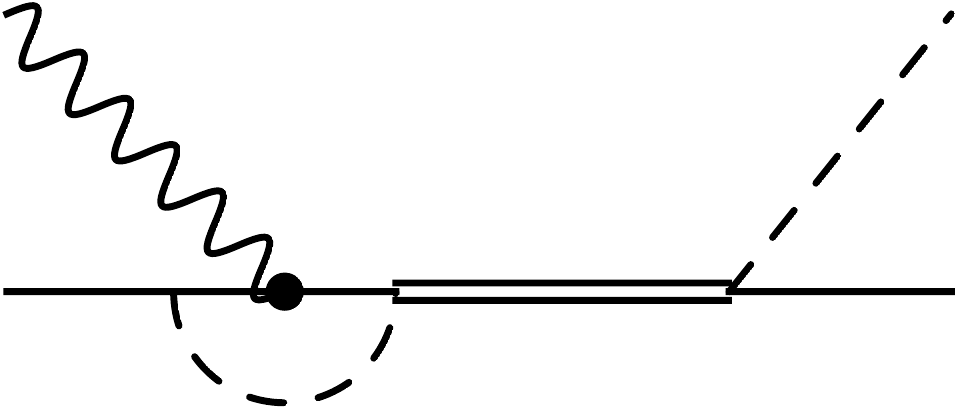}}

\caption{Feynman diagrams with a $\Delta$ propagator. Diagrams (a) and (b) shows tree graphs, that scale as $e p^{3/2}$ for $\omega_\pi \sim M_\pi$. For $\omega_\pi \sim \Delta_M$ the uncrossed diagram scales as $ep^{-1}$ and the crossed as $ep$. Diagram (c) scales as $ep^{5/2}$ when $\omega_\pi \sim M_\pi$ and $e p^0$ when $\omega_\pi \sim \Delta_M$. Diagram (d), when $\omega_\pi \sim M_\pi$, scales as $ep^3$ and for $\omega \sim \Delta_M$ it scales as $ep$.}

\label{F:deltagraphs}

\end{figure}

Having two different power counting schemes valid at different energies presents us with another choice. One approach would be to compute all the diagrams to some nominal order in one regime up to a cut-off energy and then compute a different set of contributions for the second regime, taking care to match the contributions at the boundary. This would be tedious and ignores the fact that there is a gradual shift from one power-counting scheme regime to the other, not a sudden change. Instead, we pick a set of diagrams that are $\mathcal{O}\left(ep^3\right)$ (N$^2$LO), in the threshold region and $\mathcal{O}\left(ep\right)$ (NLO), in the resonance region, see figure \ref{F:deltagraphs}.

Selecting diagrams based on their order at two different energies realises our goal of having a theory that is accurate at low energies and adequate at high energies. Furthermore, we can simplify our work by noting that irreducible diagrams that contain a $\Delta$ and a pion loop do not vary rapidly with energy, as the power counting of the propagator is unchanged in the two regimes. The effects of these diagrams have been subsumed into the LECs (which we discuss in the next subsection).
 
\begin{figure}
\centering
\subfloat{\includegraphics[width=5cm]{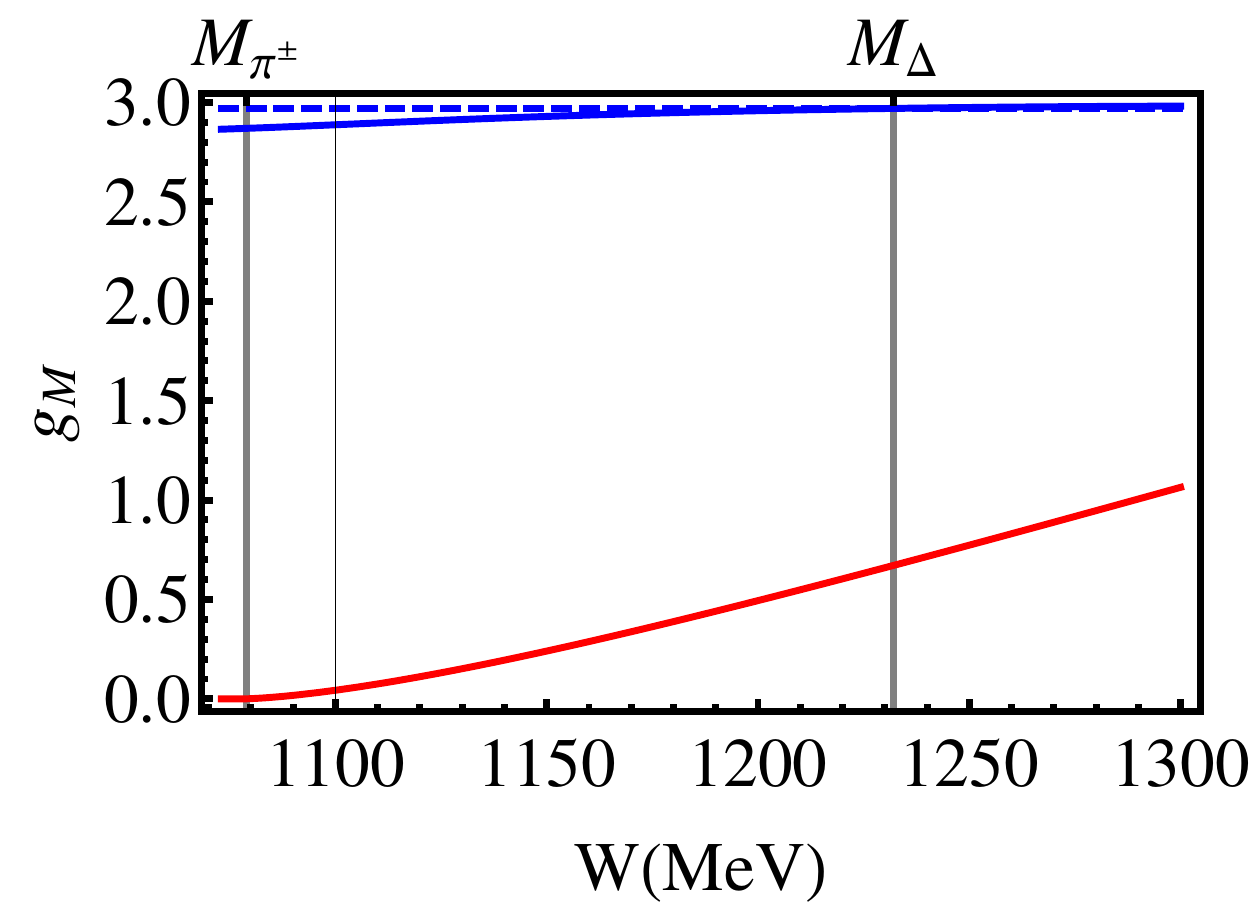}}
\subfloat{\includegraphics[width=5cm]{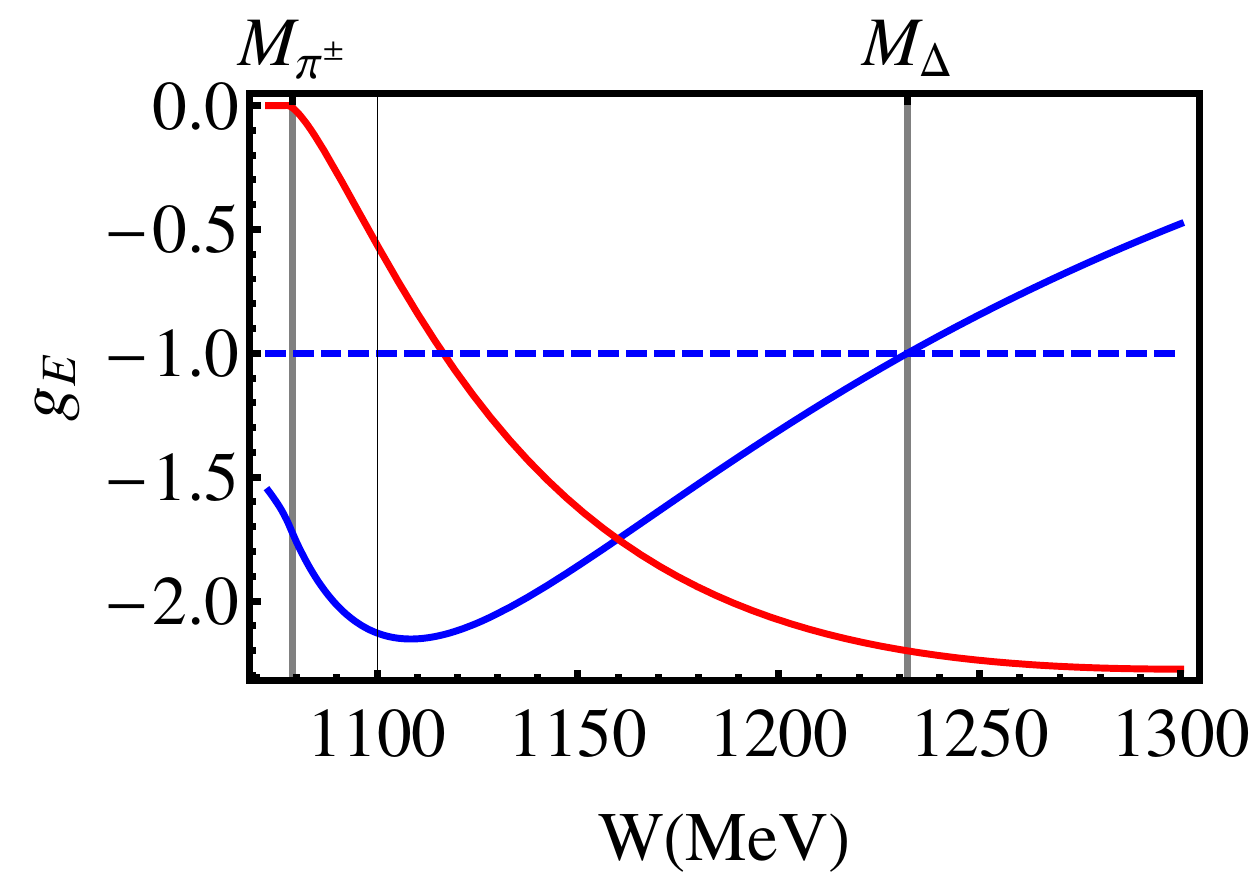}}

\caption{Plots of $g_M$ and $g_E$ in the resonance region. In these plots, at resonance, $g_M=2.97$ and $g_E=-1.0$. Blue denotes the real part of each and red the imaginary. The dashed lines show no loop effects. Note here we have corrected the sign error from \cite{mcgovern13} when presenting the results for $g_E$.}

\label{F:gplots}

\end{figure}

We can absorb the effects of the vertex corrections, diagrams (c) and (d) from figure \ref{F:deltagraphs}, into the running with energy of the electric and magnetic $\gamma$N$\Delta$ couplings, $g_M$ and $g_E$ \cite{mcgovern13}. The results are shown in figure \ref{F:gplots}.

\begin{figure}
\centering
\subfloat{\includegraphics[width=4.8cm]{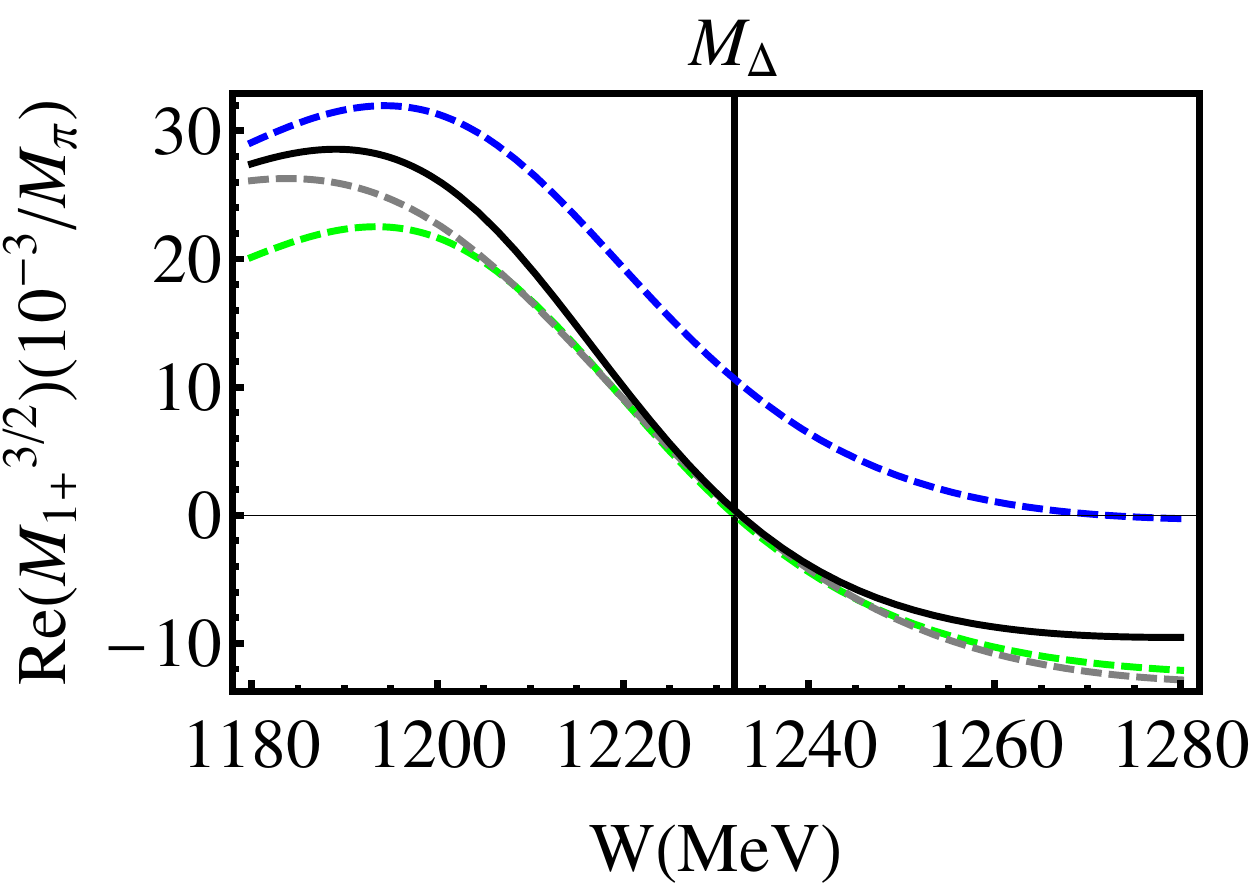}}\:
\subfloat{\includegraphics[width=4.8cm]{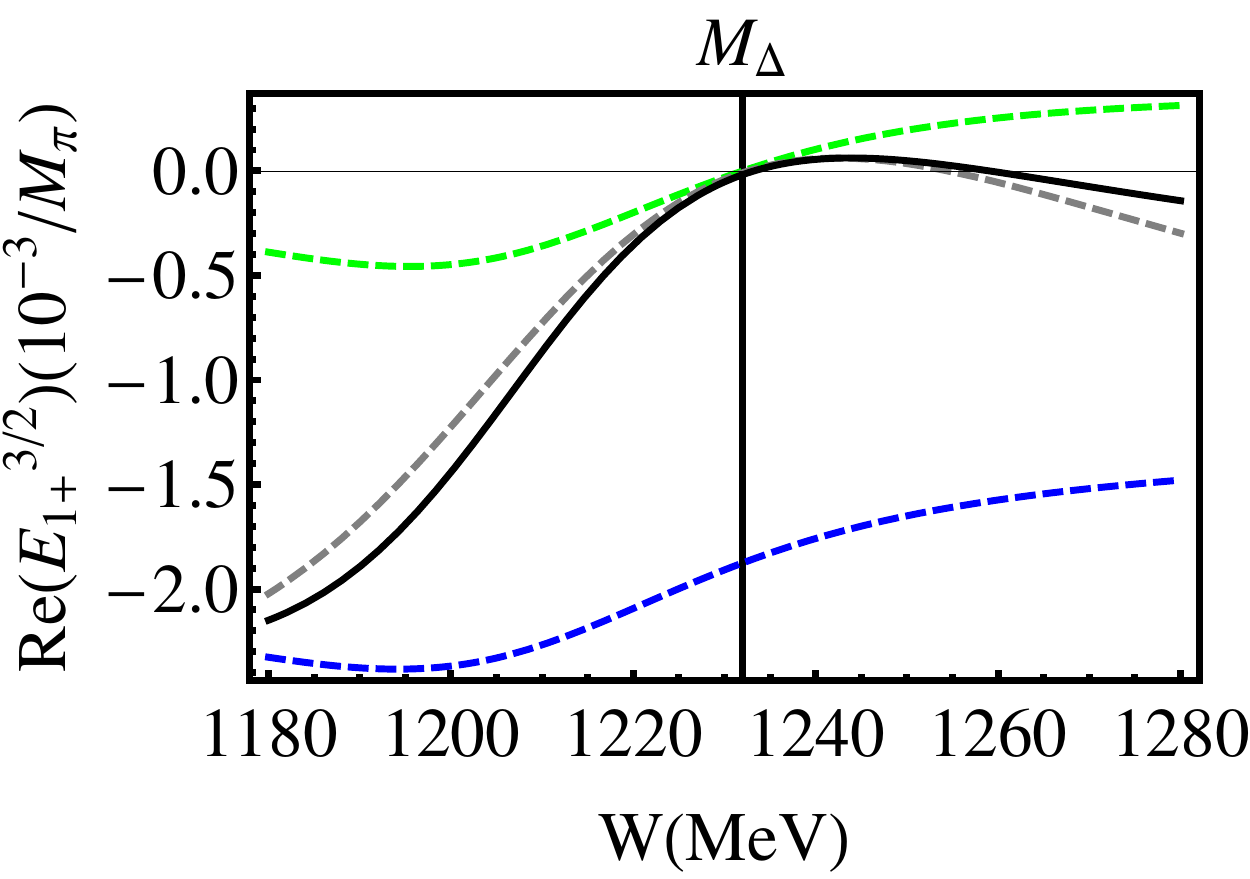}}\\
\subfloat{\includegraphics[width=4.8cm]{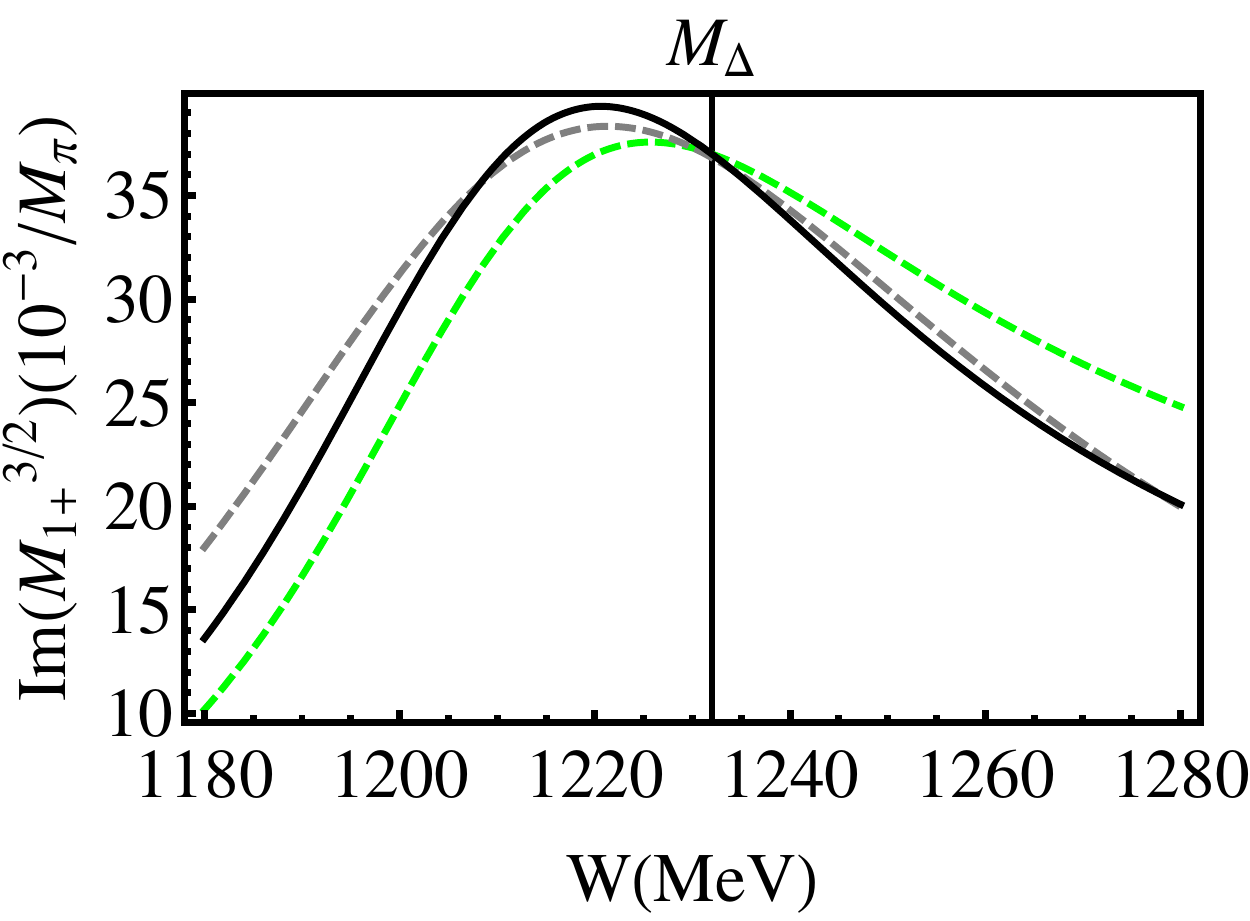}}\:
\subfloat{\includegraphics[width=4.8cm]{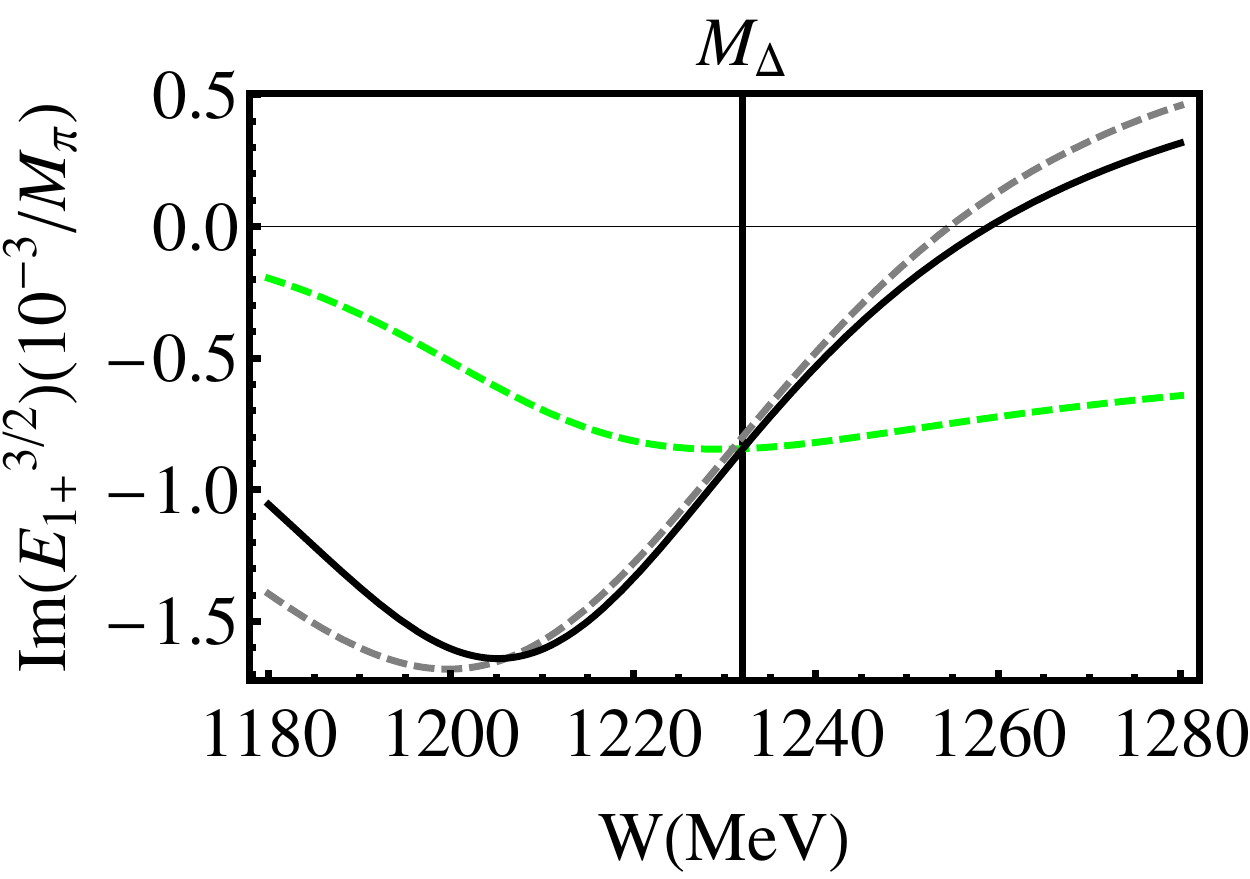}}

\caption{Plots of the $E_{1+}^{3/2}$ and $M_{1+}^{3/2}$ in the resonance region. The green dashed line is the $\Delta$ contributions only, blue dashed includes tree diagrams and solid black the vertex corrections \cite{pascalutsa06}. The dashed grey line are the results from MAID for comparison.}

\label{F:watsonplot}

\end{figure}

The vertex corrections outlined above are important to restore Watson's theorem. To remind the reader, Watson's theorem states that the real part of an amplitude in a resonant channel should be zero at resonance. The dominant channels for the $\Delta(1232)$ are the $E_{1+}^{3/2}$ and $M_{1+}^{3/2}$ partial waves. A $\Delta$-only calculation obeys Watson's theorem at resonance, including the $\pi N$ tree diagrams violates it. Including the vertex corrections restores Watson's theorem \cite{pascalutsa06}, see figure \ref{F:watsonplot}.

\subsection{Fitting procedure} \label{fitprocedure}

Working to fourth order we are presented with eight LECs that can be determined only by fitting to data. Six of these are pure photoproduction counter-terms; one, $c_4$, can be extracted from $\pi N$ scattering; and one is $g_M$, the magnetic $\gamma$N$\Delta$ coupling (we fix the electric coupling, $g_E=-1.0$) .

The contributions of the 6 photoproduction LECs to the various multipoles are as found by Hilt {\it et al.} \cite{hilt13}\footnote{This differs from what was presented in Pisa where we used the same 5 LECs as Bernard {\it et al.} \cite{bernard96,bernard01}}:

\begin{align}
 E^{ct}_{0+}(\omega_\pi) &= \frac{e\left(6{\tilde e}_{48}+2 {\tilde e}_{49}-4{\tilde e}_{50}+3 {\tilde e}_{51}\right)\omega_\pi^3}{12 \pi F}-\frac{e\left(3 {\tilde e}_{112}+{\tilde e}_{49}\right)M_\pi^2\omega_\pi}{6 \pi F}, \notag \\
 P^{ct}_1(\omega_\pi) &= -|\vec{k}_\pi|\frac{e \left(2 {\tilde e}_{48}+{\tilde e}_{51}\right)\omega_\pi^2}{4 \pi F}, \quad
 P^{ct}_2(\omega_\pi) = |\vec{k}_\pi|\frac{e {\tilde e}_{48}\omega_\pi^2}{2 \pi F}, \notag \\
 P^{ct}_3(\omega_\pi) &=-|\vec{k}_\pi|\frac{e {\tilde d}_9 \omega_\pi}{\pi F}, \quad
E^{ct}_{2-}(\omega_\pi)=-|\vec{k}_\pi|^2\frac{e {\tilde e}_{49}\omega_\pi}{6 \pi F}.
\label{eq:lecs}
\end{align}
The parameters above correspond to the following combinations of LECs:

\begin{align}
{\tilde d}_9 = d_8+d_9 ,\phantom{quad}{\tilde e}_{48} &= e_{48}+e_{67}, \phantom{quad} {\tilde e}_{49} = e_{49}+e_{68}, \notag \\
{\tilde e}_{50} = e_{50}+e_{69},\phantom{quad} {\tilde e}_{51} &= e_{51}+e_{71},\phantom{quad}{\tilde e}_{112} = e_{112}+e_{113},
\label{eq:laglecs}
\end{align}
in terms of the LECs from the Lagrangian in \cite{fettes00}.

Some of the diagrams contributing to fourth-order photoproduction have a second-order $\pi N \rightarrow \pi N$ vertex, see diagrams (e) and (f) in figure \ref{F:order4d}. This gives rise to a dependence on the LEC $c_4$ in our analysis\footnote{The work presented in Pisa used a value for $c_4$ obtained from third order $\pi N$ scattering. We believe this to be the reason to why we needed to adjust $g_M$ so dramatically to obtain reasonable results.}. To obtain a value for $c_4$ in-line with our work we have to fit the theory at the appropriate order to $\pi N$ data. We have done this using a second-order HB$\chi$PT calculation \cite{fettes98} combined with the relativistic Born $\Delta$ contributions \cite{chen13}. We have chosen to fit this to the real parts of the WI08 data from SAID for the S and P partial waves. The imaginary parts vanish at second-order. We find $c_4=1.18$ GeV$^{-1}$.

\begin{figure}
\centering
\subfloat{\includegraphics[width=4.8cm]{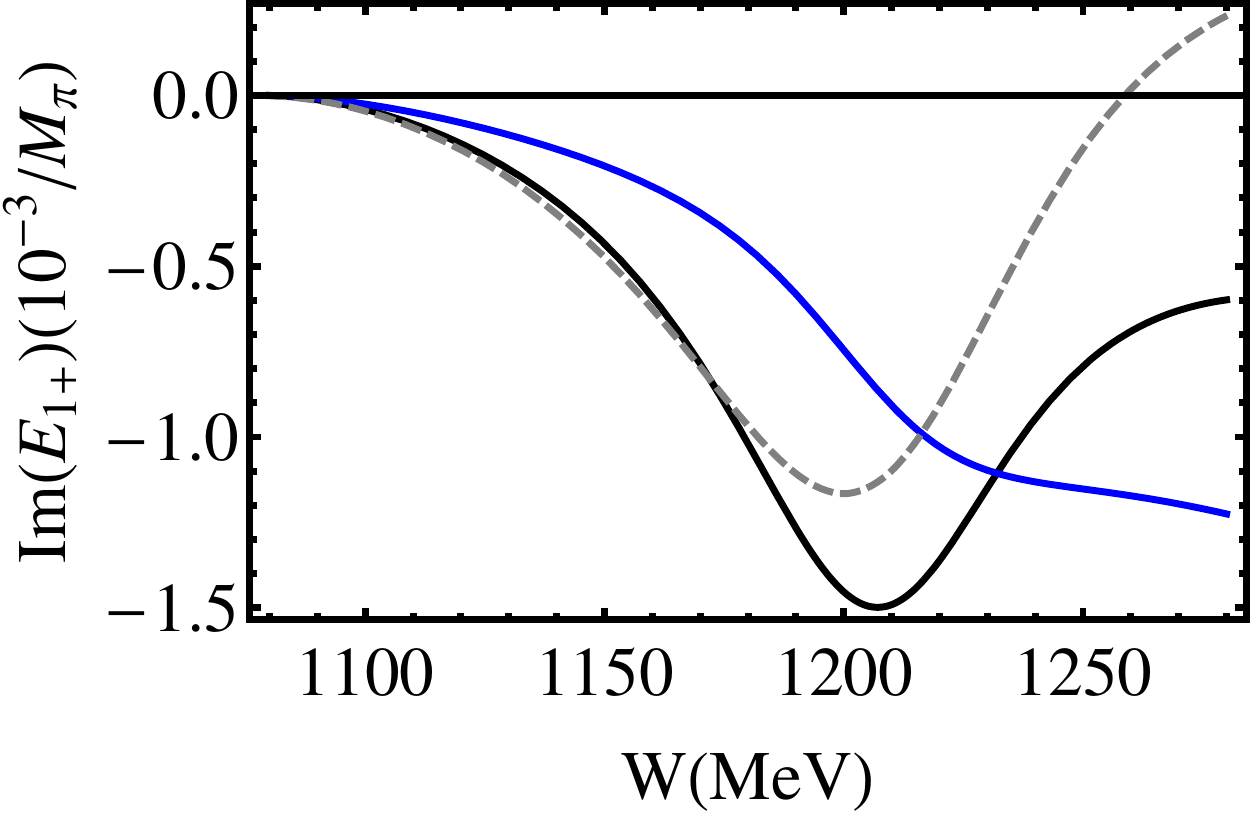}}\;
\subfloat{\includegraphics[width=4.8cm]{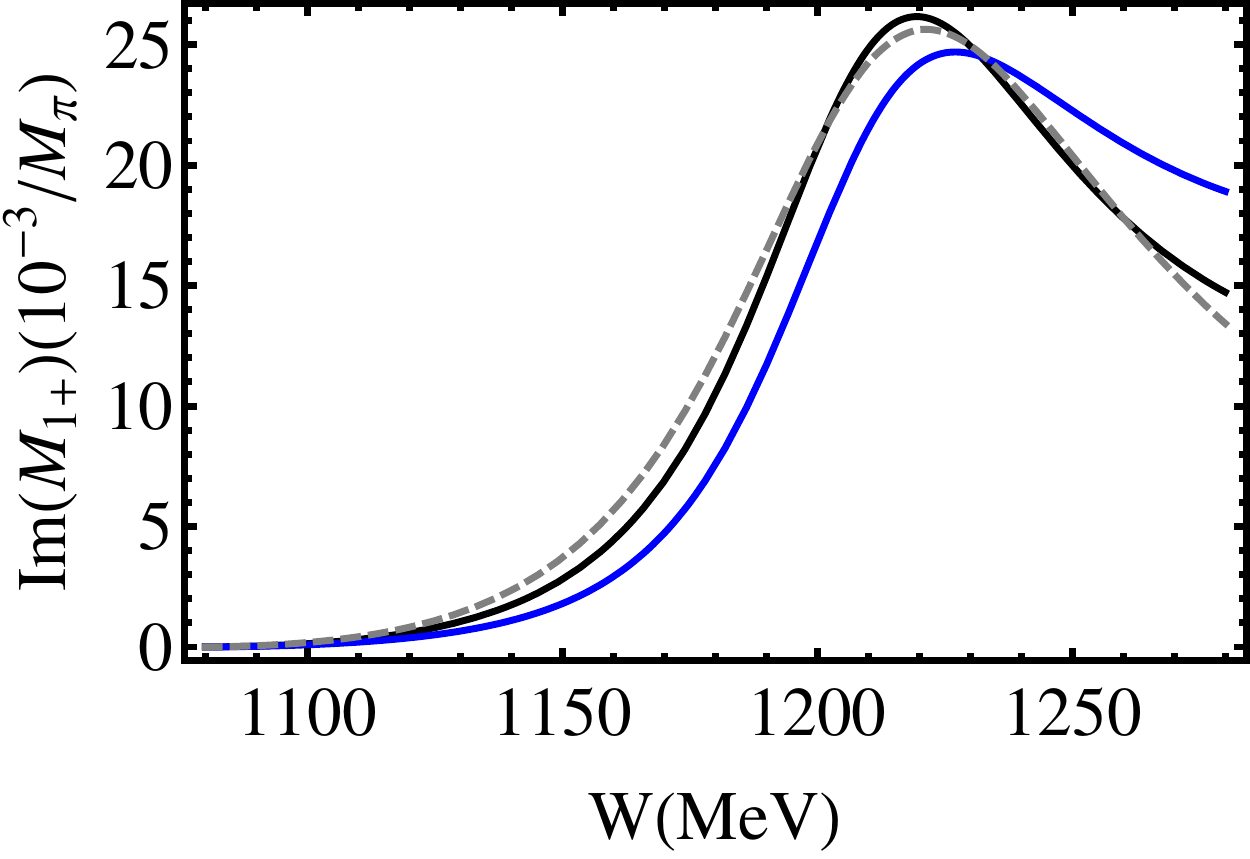}}\;
\subfloat{\includegraphics[width=4.8cm]{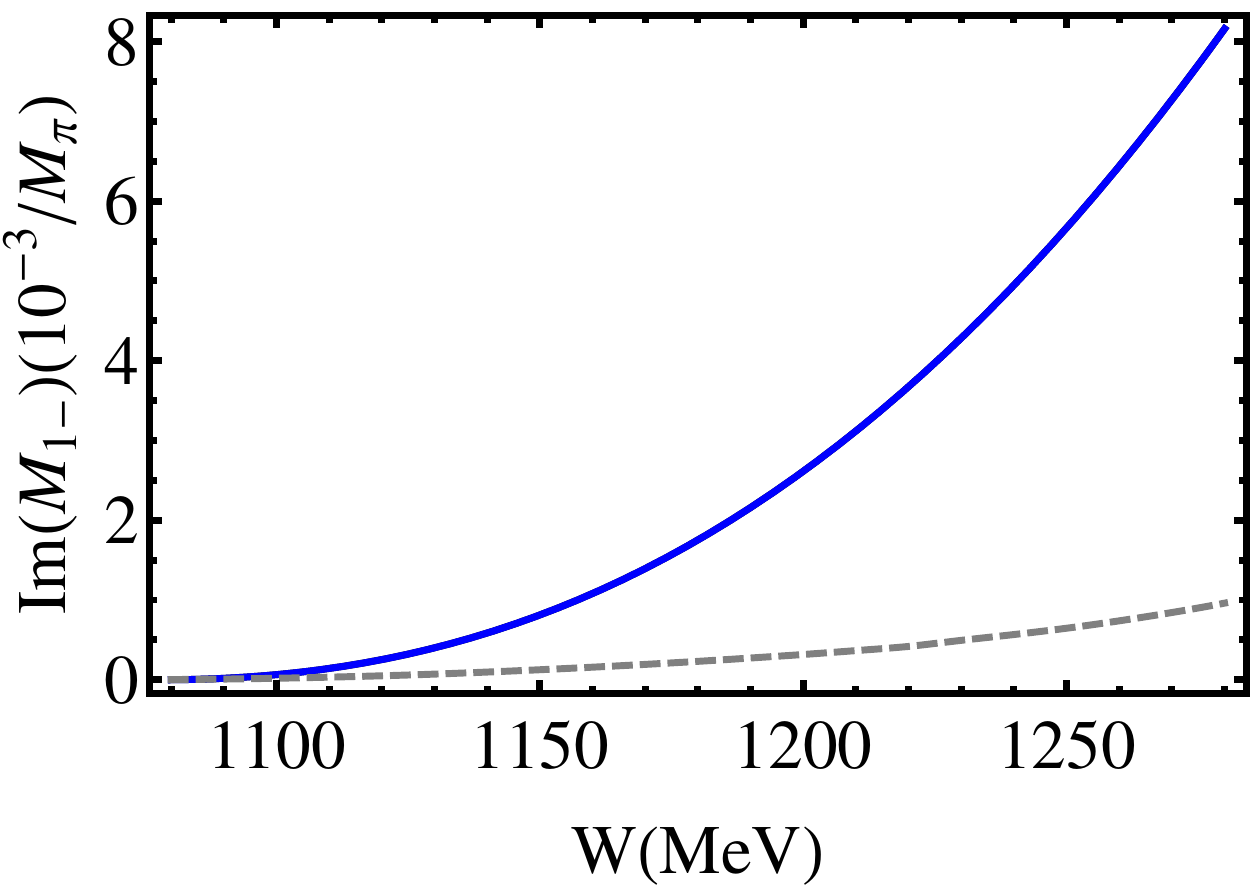}}

\caption{Imaginary parts of P-wave multipoles fitted to MAID 2007 giving $g_M=2.66$.The black line includes vertex corrections to $\Delta$ diagrams, the blue does not and the dashed grey is the result from MAID.}

\label{F:immultplot}

\end{figure}

Finally, as the original extraction of $g_M=2.9$ by Pascalutsa {\it et al.} \cite{pascalutsa06} did not take into account the effects of pion loops we must re-fit it. Furthermore, there have been studies on Compton scattering that suggest that it should be reduced by 10\% \cite{mcgovern13}. To find $g_M$ we have performed a fit to the imaginary parts of the P-wave multipoles as listed on MAID 2007. We have chosen to do this as the imaginary parts of partial waves are well known in dispersion theory and they are independent of the six photoproduction LECs (see equation \ref{eq:lecs}). The result of this fit is $g_M=2.66$, the graphs of which can be found in figure \ref{F:immultplot}.

\section{Results}

We have performed a fit to minimise $\chi^2$ by varying the 6 LECs (see equation \ref{eq:lecs}) from $E_\gamma=154$ MeV to $E_\gamma=350$ MeV for both the differential cross section, $d\sigma/d\Omega$, and the photon asymmetry, $\Sigma$. For the later observable we have experimental data only from $E_\gamma=154$ MeV to $E_\gamma=206$ MeV and $E_\gamma =250$ MeV to $E_\gamma=317$ MeV \cite{hornidge13}. We also include the systematic errors, of 4\% and 5\% for $d\sigma/d\Omega$ and $\Sigma$ respectively, in our fitting procedure. We do not compare to data beyond $E_\gamma=350$ MeV as we expect the expansion to break down for $p>0.5$.

To assess the importance of D-waves we took three variations of our calculations: truncated to P-waves, truncated to D-waves, and no truncation (i.e., including all partial waves). So far we are yet to see any significant change between the three calculations so it appears that D-waves do not play a significant role. This might change with data on other observables more sensitive to higher order partial waves \cite{fernandez-ramirez09}.

\begin{figure}
\centering
\subfloat[]{\includegraphics[height=3.8cm]{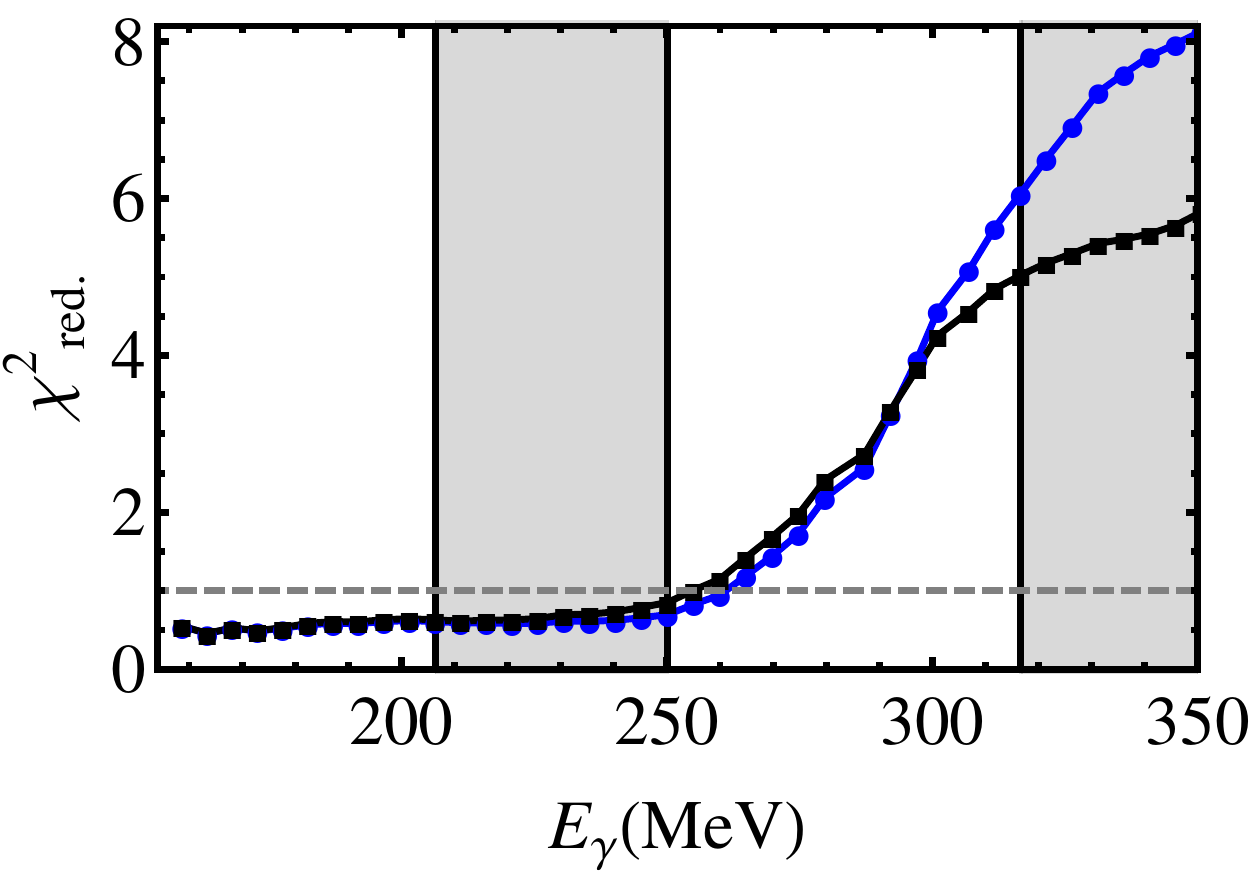}}\;
\subfloat[]{\includegraphics[height=3.8cm]{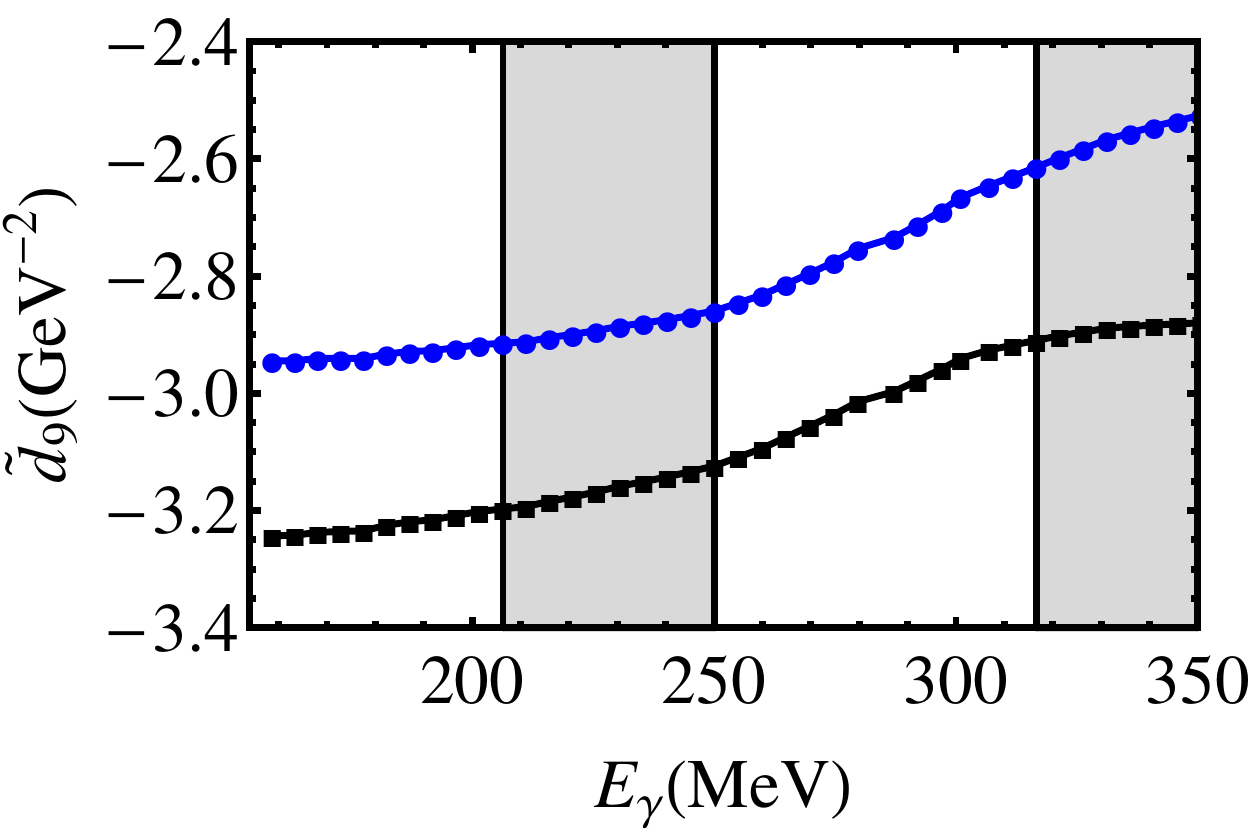}}
\caption{Plots of the $\chi^2_{red.}$ and LEC $\tilde{d}_9$ against maximum photon energy fitted. The shaded regions show where there is no data for $\Sigma$. The black line includes vertex corrections, blue line does not. The data used for each fit starts at $E_\gamma=154$ MeV and ends at each point on the plot.}
\label{F:chilec26plot}
\end{figure}

We proceed in our analysis by fitting to data including the vertex corrections. A plot of the reduced $\chi^2$ against the photon energy can be found in figure \ref{F:chilec26plot} (a); we fit from $E_\gamma=154$ MeV up to a maximum photon energy denoted by each plot marker on the graph. From these results we can see that, close to threshold, the calculations with and without vertex corrections are indistinguishable and are accurate up-to $E_\gamma \approx 260$ MeV. Beyond $E_\gamma \approx 290$ MeV we see how the inclusion of the vertex corrections improves the fit. 

\begin{figure}
\centering
\subfloat{\includegraphics[width=5cm]{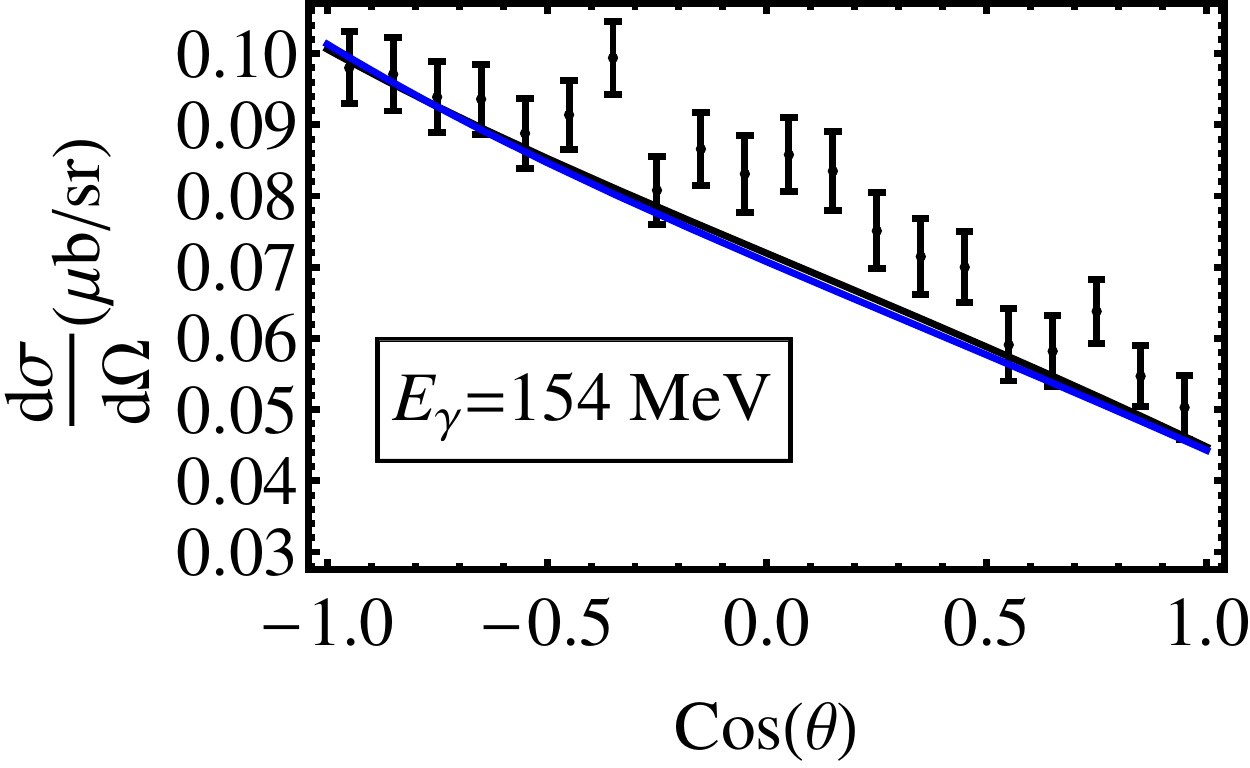}}\:
\subfloat{\includegraphics[width=5cm]{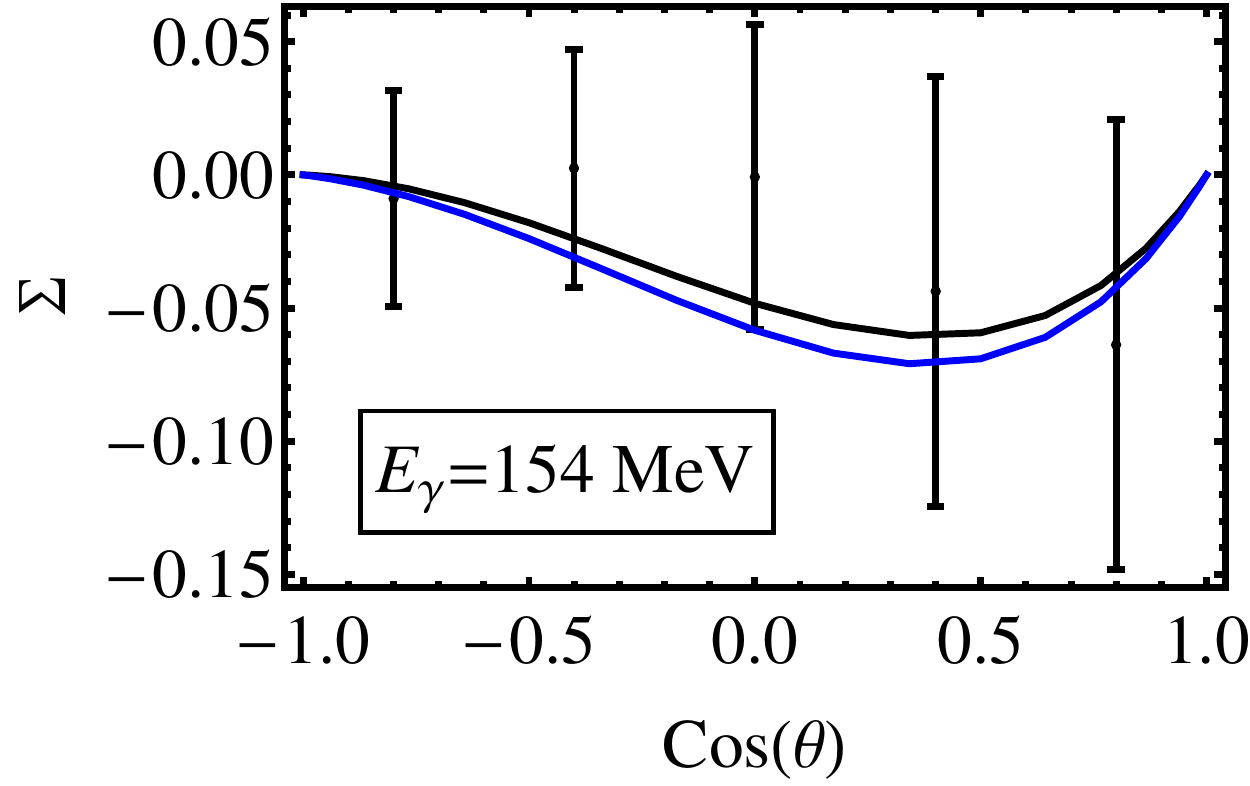}}\\
\subfloat{\includegraphics[width=5cm]{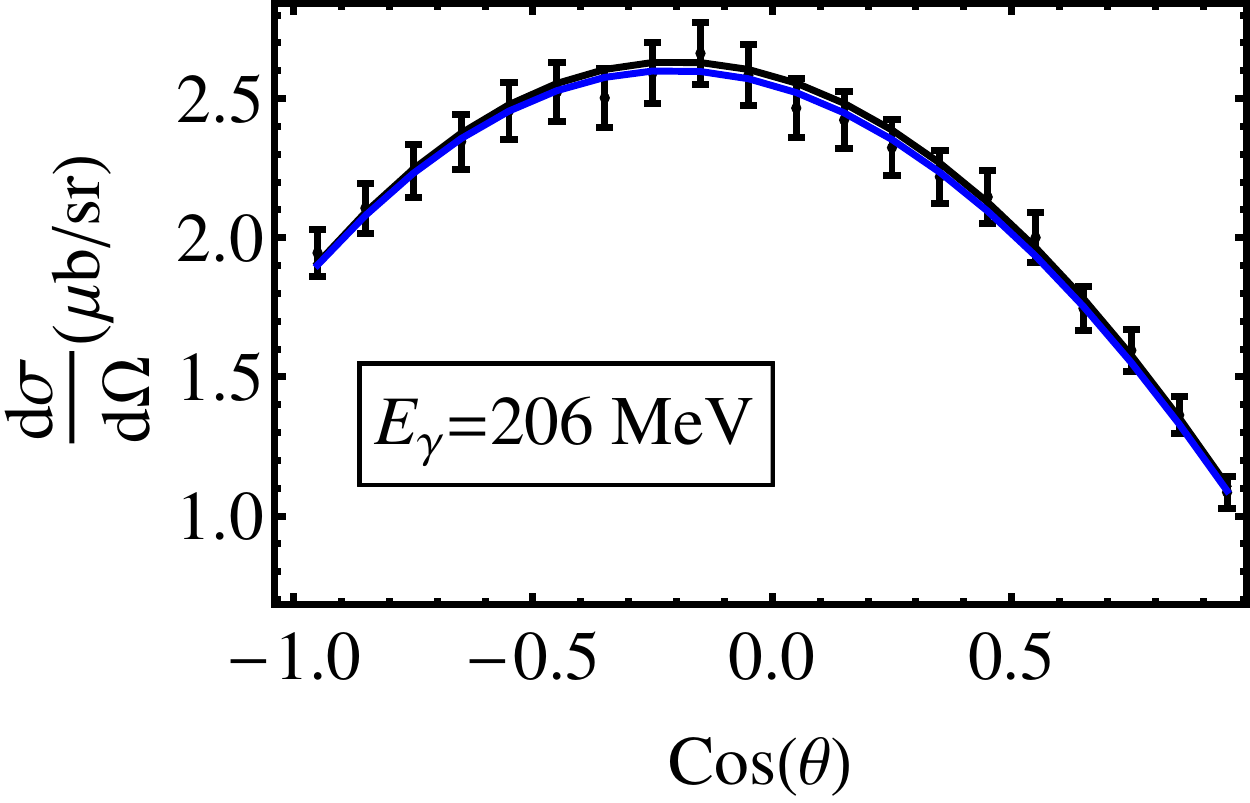}}\:
\subfloat{\includegraphics[width=5cm]{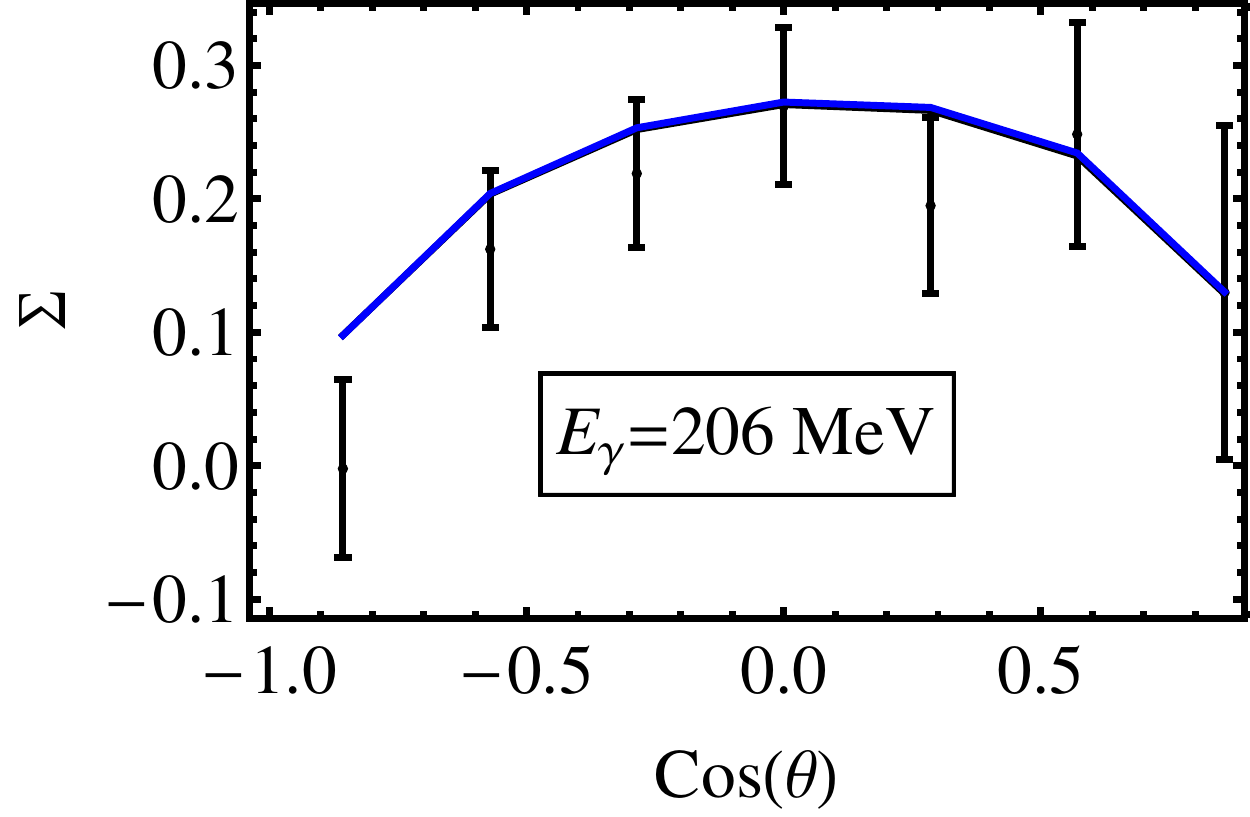}}\\
\subfloat{\includegraphics[width=5cm]{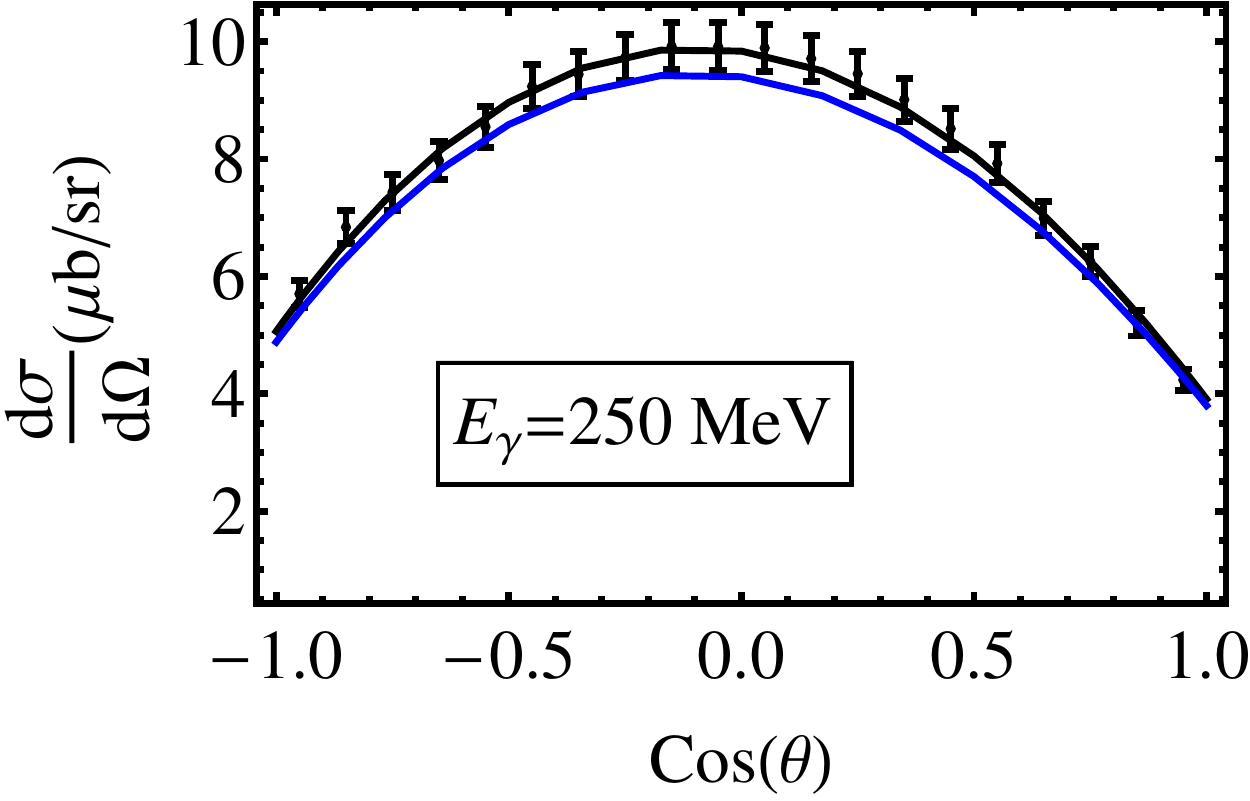}}\:
\subfloat{\includegraphics[width=5cm]{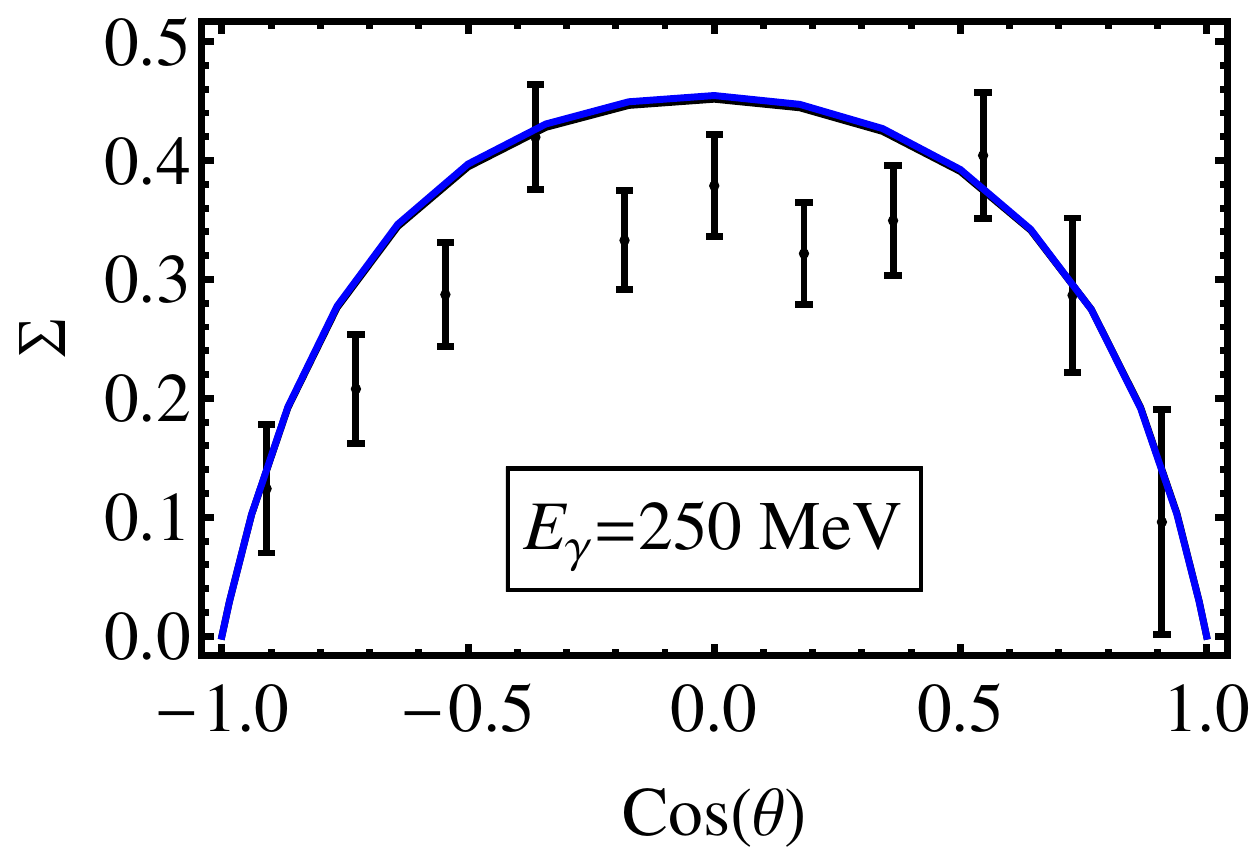}}\\
\subfloat{\includegraphics[width=5cm]{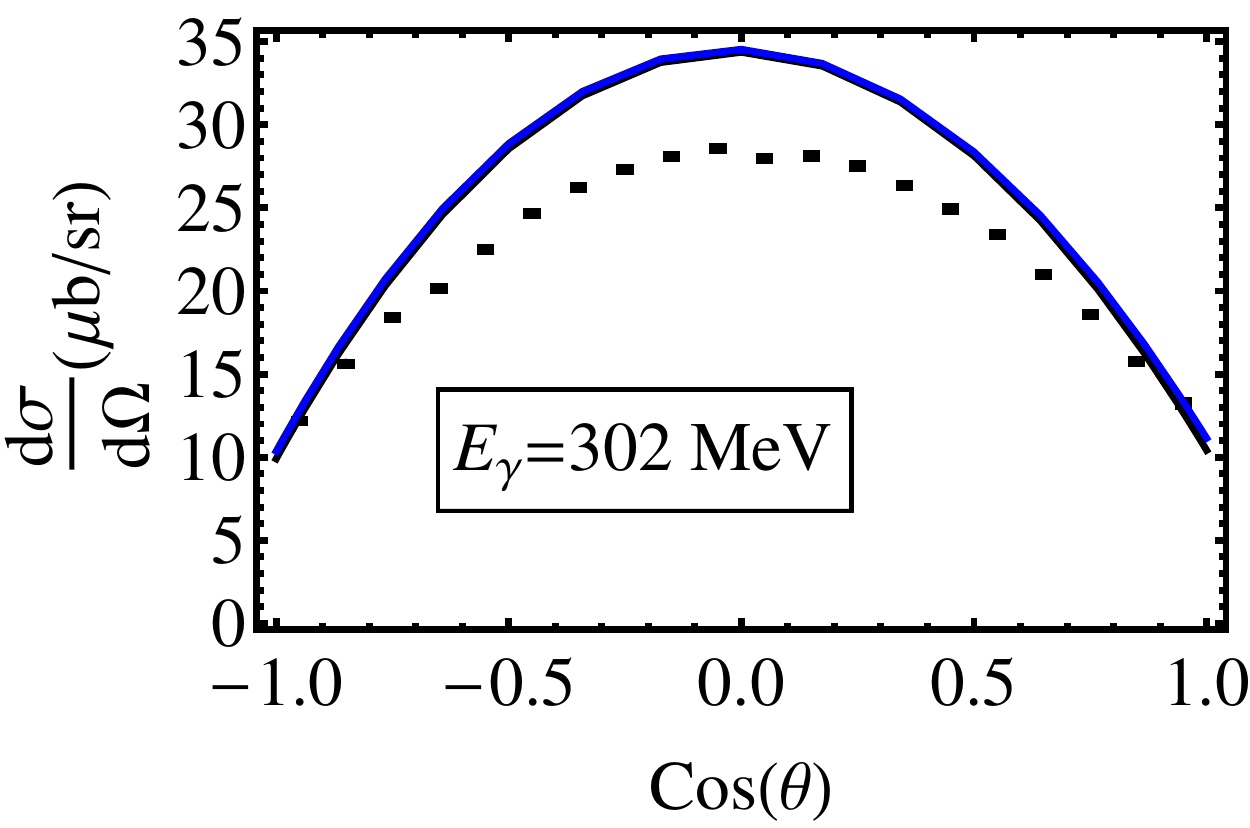}}\:
\subfloat{\includegraphics[width=5cm]{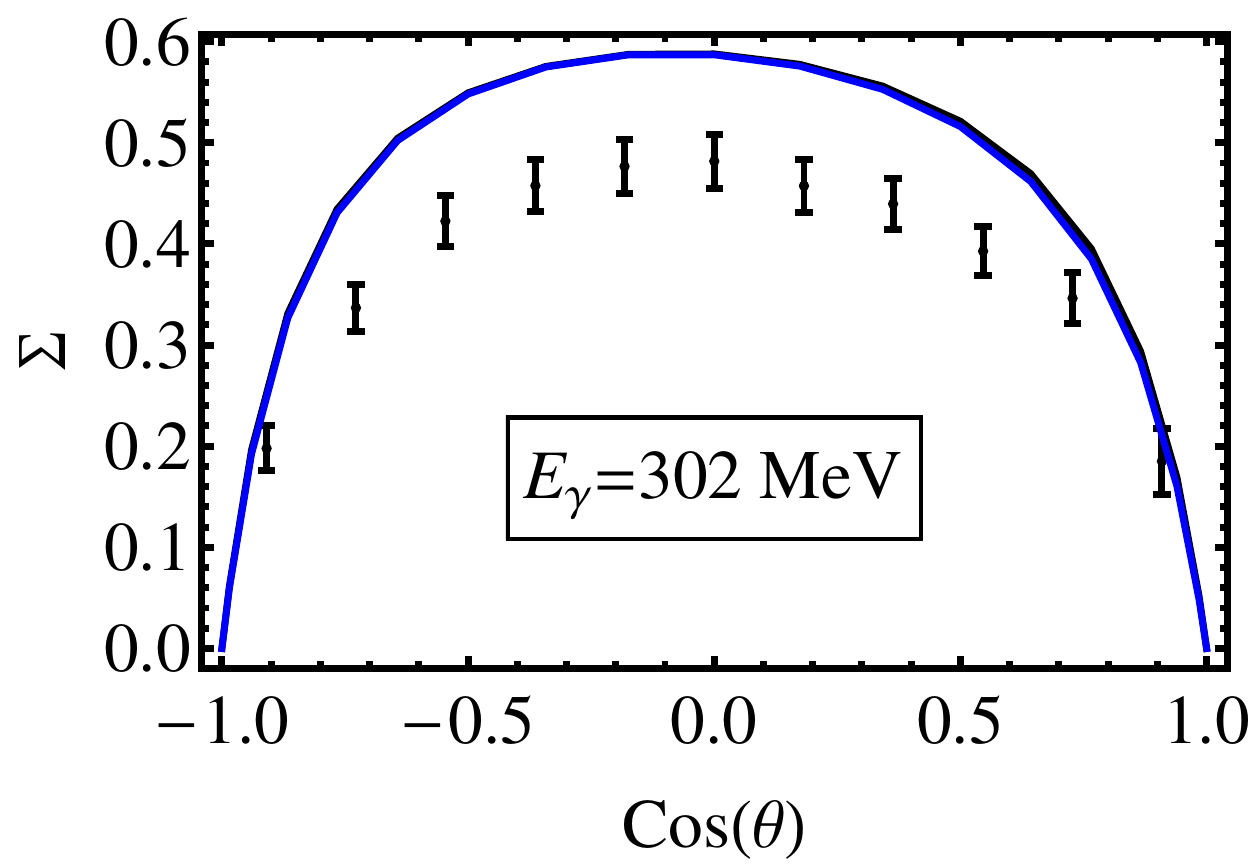}}\\
\subfloat{\includegraphics[width=5cm]{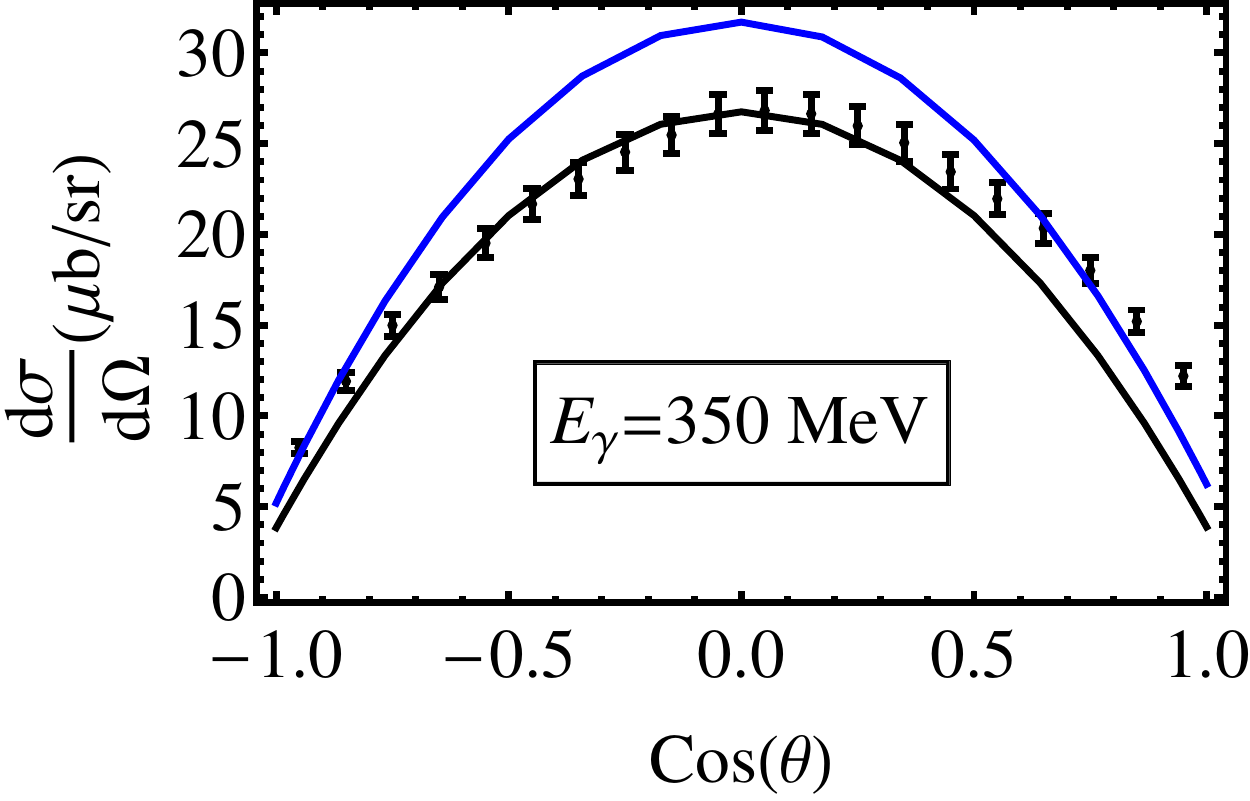}}

\caption{Various plots of data compared to our calculations. The black line includes vertex corrections to $\Delta$ diagrams, the blue line does not. These plots use the LECs after fitting from $E_\gamma=154$ MeV to $E_\gamma=350$ MeV.}

\label{F:data26plot}

\end{figure}

Plotting both $d\sigma/d\Omega$ and $\Sigma$ at various energies we can see how our calculations compare to the experimental data, see figure \ref{F:data26plot}. Both versions of our calculations appear to overestimate the data above $E_\gamma\approx260$ MeV. At energies beyond $E_\gamma \approx 310$ MeV, including the vertex corrections appears to improve the fit; this is reflected in the change of slope in figure \ref{F:chilec26plot} (a).

A test to see if the fit has been successful is to examine the energy dependence of the LECs in eq.\ref{eq:laglecs}. There does appear to be some energy dependence in the parameters. We have shown the variation of $\tilde{d}_9$, the only LEC that enters at $\mathcal{O}(p^3)$, against range of energy fitted in \ref{F:chilec26plot} (b). As this is still a work in progress, we are yet to test if the variations can be contained within a $\chi^2_{red.}+1$ error-band of the fit as explored, without the inclusion of the $\Delta$, in \cite{fernandez-ramirez13}.

\begin{figure}
\centering
\subfloat{\includegraphics[width=4.8cm]{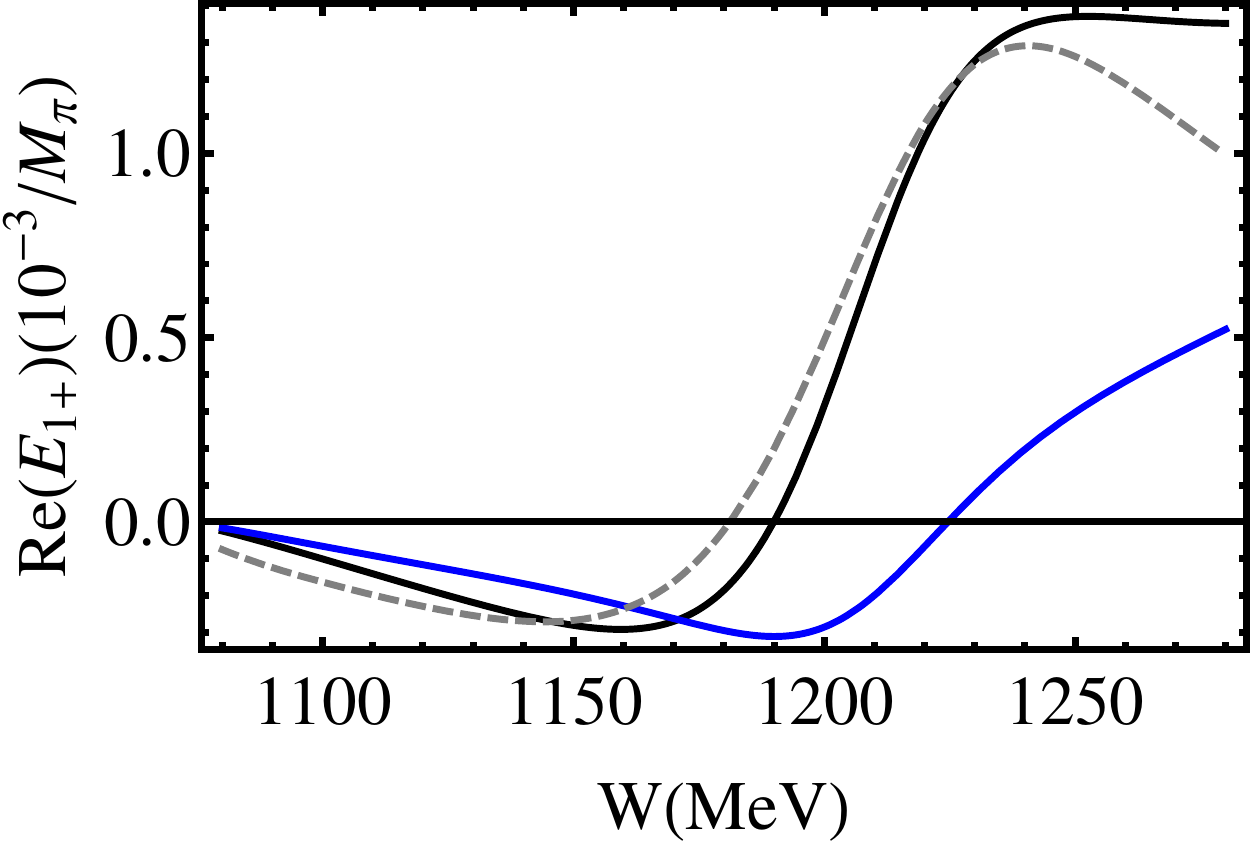}}\;
\subfloat{\includegraphics[width=4.8cm]{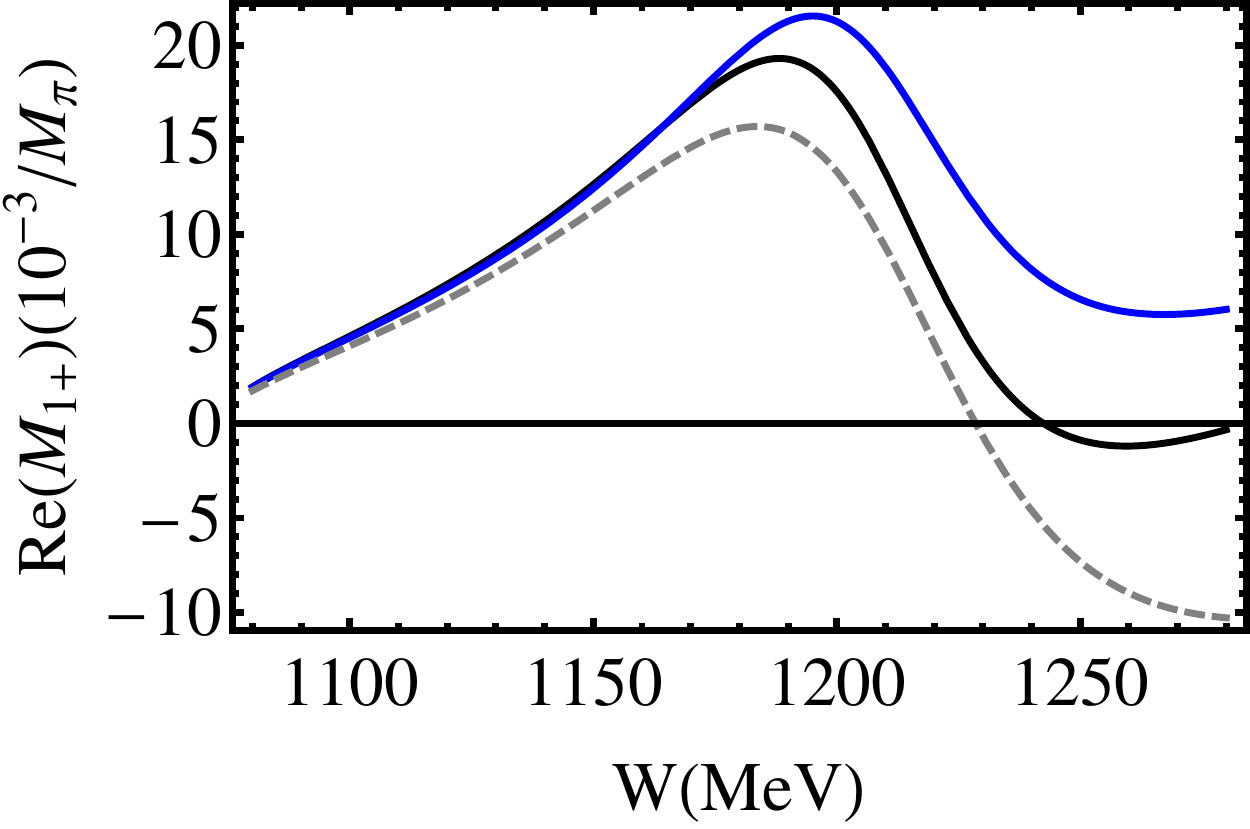}}\;
\subfloat{\includegraphics[width=4.8cm]{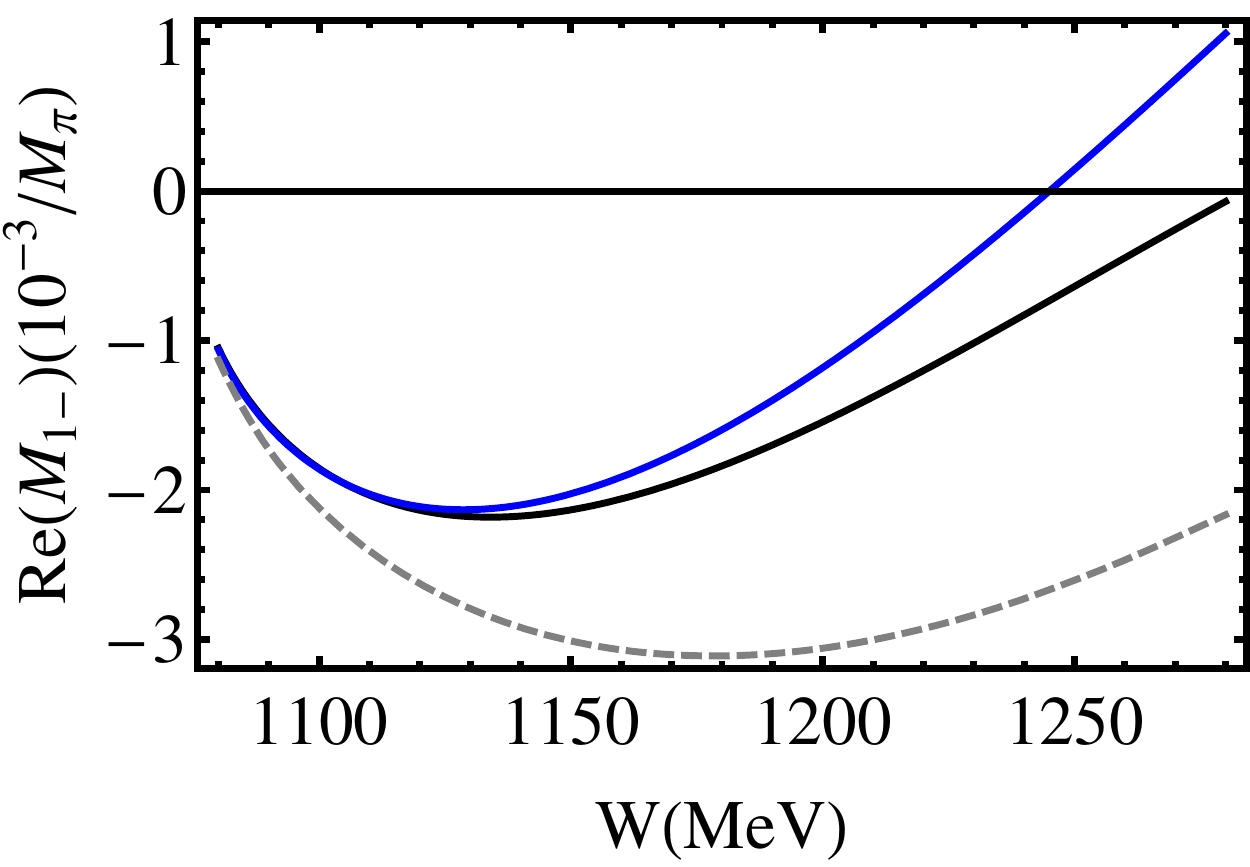}}

\caption{Real parts P-wave multipoles for our calculations to complement the imaginary plots above. The black line includes vertex corrections to $\Delta$ diagrams, the blue line does not and the dashed grey is the result from MAID.}

\label{F:remultplot}

\end{figure}

For completeness we also show in figure \ref{F:remultplot}  the real parts of the P-wave partial waves, to compliment the imaginary parts shown earlier in figure \ref{F:immultplot}.

A point worth noting is that if $c_4$ is reduced to a value of $-$2.17 GeV$^{-1}$ then $g_M$ can be increased to 2.8, and we can re-create the imaginary parts of the P-waves as found by MAID almost exactly. Reducing $c_4$ can also reduce the already small energy dependence of the LECs mentioned above. But a value of $c_4=-2.17$ GeV$^{-1}$ is far from what can be justified by $\pi N$ scattering.

\section{Conclusions}

We have embarked on a comparison between the predictions of chiral EFTs for neutral pion photoproduction and recent data \cite{hornidge13} from threshold to energies approaching the $\Delta(1232)$ resonance. This is a work in progress but preliminary results are encouraging as our results show a clear improvement from the $\Delta$-less $\mathcal{O}(p^4)$ heavy baryon and relativistic results published by Fern\'andez-Ram\'irez {\it et al.} and Hilt {\it et al.} respectivley \cite{fernandez-ramirez13,hilt13}.

\end{document}